\documentclass[%
reprint,
superscriptaddress,
showpacs,
nofootinbib, nobibnotes, %footnote in the first page instead of the bib
 amsmath,
 amssymb,
 aps,
 prl
]{revtex4-1}

\usepackage{microtype}% improves general appearance of the text
\usepackage{blindtext}
\usepackage{dcolumn}% Align table columns on decimal point
\usepackage{bm}% bold math
\usepackage{color, soul, colortbl} %to highlight text ==> \hl{text}
\usepackage[colorlinks=true,
						linkcolor=blue,
						urlcolor=blue,
						citecolor=blue,
						bookmarks=true,
						pdfborder={0 0 0}]{hyperref}% add hypertext capabilities

\usepackage{xcolor}
\usepackage{mdframed} %box around text         

\usepackage{graphicx,epsfig}% Include figure files
\usepackage{subfigure}
\usepackage{amsmath,amsfonts}%,multirow,rotate,color}amssymb
\usepackage{xfrac}%simple fraction layout \sfrac{}{}
\usepackage{enumerate}
\usepackage[overload]{empheq}
\usepackage{mathtools}
\usepackage{cases} 
\usepackage{cancel} % to make a strikethrough in math mode (diagonal bar)

\begin{document}
\title{Revealing the determinants of gender inequality in urban cycling with large-scale data}

\author{Alice Battiston}
\affiliation{University of Turin, Via Giuseppe Verdi, 8, 10124 Torino TO, Italy}
\author{Ludovico Napoli}
\affiliation{Central European University, Quellenstraße 51, 1100 Wien, Austria}
\author{Paolo Bajardi}
\affiliation{ISI Foundation, Via Chisola 5, 10126 Torino TO, Italy}
\author{André Panisson}
\affiliation{ISI Foundation, Via Chisola 5, 10126 Torino TO, Italy}
\author{Alan Perotti}
\affiliation{ISI Foundation, Via Chisola 5, 10126 Torino TO, Italy}
\author{Michael Szell}
\affiliation{IT University of Copenhagen, Rued Langgaards Vej 7, 2300 København, Denmark}
\affiliation{ISI Foundation, Via Chisola 5, 10126 Torino TO, Italy}
\affiliation{Complexity Science Hub, Josefstädter Str. 39, 1080 Wien, Austria}
\author{Rossano Schifanella}
\affiliation{University of Turin, Via Giuseppe Verdi, 8, 10124 Torino TO, Italy}
\affiliation{ISI Foundation, Via Chisola 5, 10126 Torino TO, Italy}

%\date{\today}

%%%%%%%%%%%%%%%%%%%%%%%%%%%%%%%%%%%%%%%%%%%%%%%%%%%%%%%%%%%%%%%%
\begin{abstract}
Cycling is an outdoor activity with massive health benefits, and an effective solution towards sustainable urban transport. Despite these benefits and the recent rising popularity of cycling, most countries still have a negligible uptake. This uptake is especially low for women: there is a largely unexplained, persistent gender gap in cycling.
To understand the determinants of this gender gap in cycling at scale, here we use massive, automatically-collected data from the tracking application Strava on outdoor cycling for 61 cities across the United States, the United Kingdom, Italy and the Benelux area. Leveraging the associated gender and usage information, we first quantify the emerging gender gap in recreational cycling at city-level. A comparison of cycling rates of women across cities within similar geographical areas unveils a broad range of gender gaps. On a macroscopic level, we link this heterogeneity to a variety of urban indicators and provide evidence for traditional hypotheses on the determinants of the gender-cycling-gap. We find a positive association between female cycling rate and urban road safety. On a microscopic level, we identify female preferences for street-specific features in the city of New York. Enhancing the quality of the dedicated cycling infrastructure may be a way to make urban environments more accessible for women, thereby making urban transport more sustainable for everyone.
\end{abstract}

%%%%%%%%%%%%%%%%%%%%%%%%%%%%%%%%%%%%%%%%%%%%%%%%%%%%%%%%%%%%%%%%
\maketitle

\begin{figure*}[htbp]
\centering
\includegraphics[width=1\linewidth]{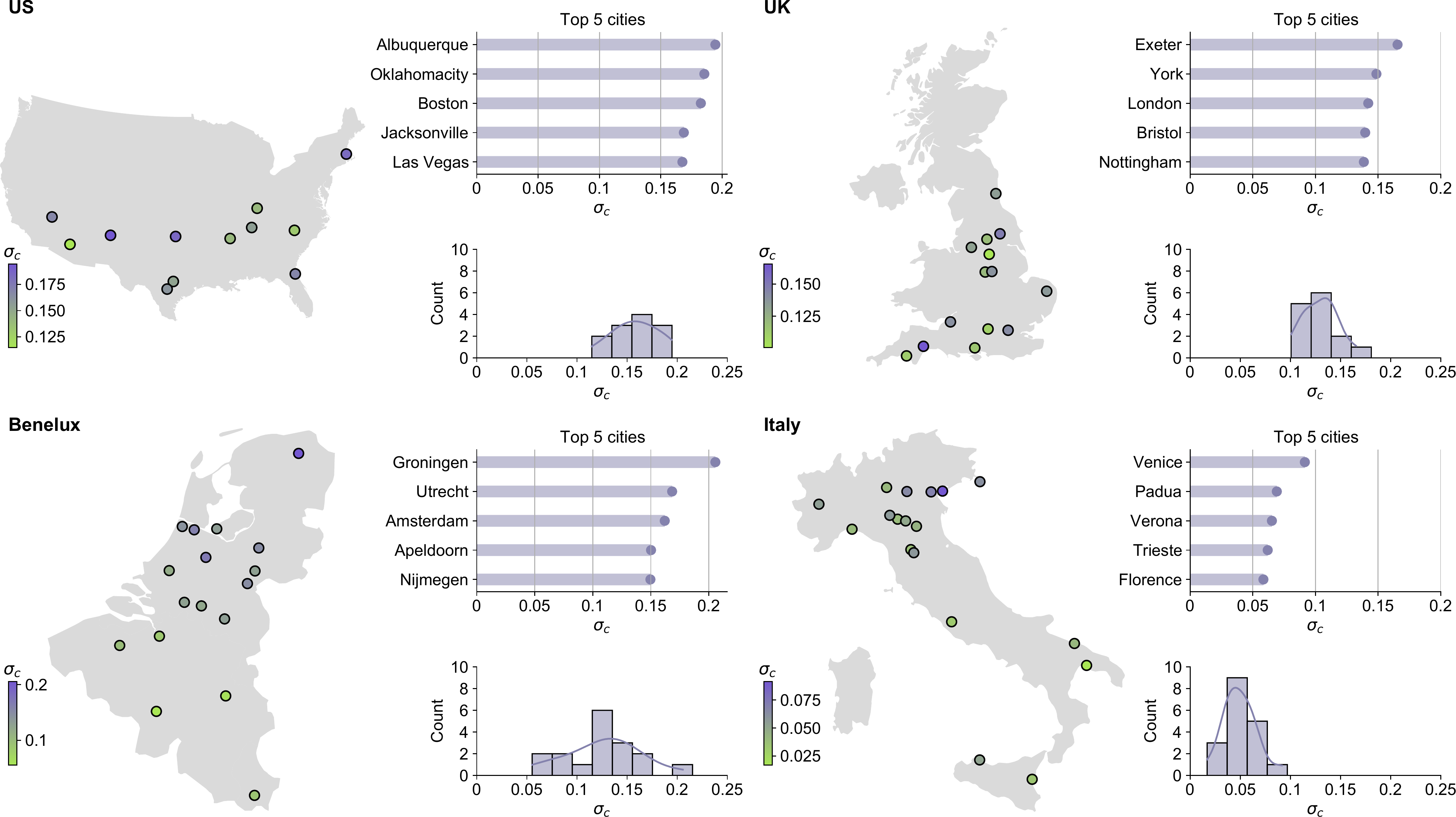}
\caption{\textbf{Gender gap in recreational cycling in Strava: overview of cities included in the study}. For each of the four geographical areas covered by the study, the figure depicts: the location of cities included in the analysis, the value of the female cycling rate $\sigma_c$ for the five cities displaying the highest $\sigma_c$ and the distribution of $\sigma_c$ in the geographical area.}
\label{fig:citiesoverview}
\end{figure*}

Cycling is an outdoor activity associated with many individual and societal benefits. From the individual perspective, cycling has a positive impact on both physical and mental health, with a strong link to improved cardio-respiratory fitness, decreased cardiovascular mortality risk, and reduced stress-levels \cite{oja2011health, leyland2019effect, avila2017relationship}. From a societal viewpoint, cycling is an environmentally-friendly and highly economic commuting option, especially for typical urban trips \cite{gossling2019sca}. Recently, the United Nation (UN) Sustainable Development Goals (SDG) identified it as a pivotal component of a sustainable urban-mobility system \cite{SDG}. Interventions targeted at increasing the number of cyclists are recommended as one of the solutions against traffic congestion, increased emissions, poor air quality and road safety. 

Despite these wide-ranging benefits, cycling is mostly a male-dominated activity with a large gap in participation rates between men and women. Cycling research and policy making that is mostly focused on improving mobility for the existing, dominant group, risks to ignore half of the population and sustainable mobility solutions for everybody \cite{monk1982not,hanson2010gmn}. Data on the use of bike-sharing services in three large US cities (New York, Boston and Chicago) show that only one in four bicycle trips in the 4-year period between 2014 and 2018 was made by a woman; other modes of transport, however, do not display comparable trip-share gaps \cite{hosford2019quantifying}. Similarly, in San Francisco, only 29\% of cyclists are women \cite{funaki2019don}. Recent data for England show that on average, not only do men take more bicycle trips per week than women, but they also cover longer distances \cite{Ukcycling}. A few European countries however, such as Denmark, Germany and the Netherlands represent the main exception to this pattern, with women making up for more than 45\% of all cyclists in these areas already in 2005 \cite{pucher2008making}. 
The evidence from this group of countries demonstrates that the reason for any kind of gender gap is not intrinsic but comes from place-specific barriers that need to be identified and, whenever possible, removed, if cycling should become a universal mode of transport. 

The academic literature aimed at understanding the determinants of the gender gap in cycling links it on one hand to behavioral and psychological hypotheses. Women perceive cycling as a riskier activity compared to men, which would directly translate into a stronger preference for cycling infrastructure that is physically separated from motorized traffic \cite{aldred2017cycling, garrard2008promoting, misra2018modeling, dill2014can}. On the other hand, physical route characteristics can play a role, for example in San Francisco, where women disfavor steep slopes, particularly for commuting \cite{hood2011gps}. In low-cycling contexts, women also report other deterring factors, such as an aversion for long distances and poor weather conditions, and a generally lower confidence in their cycling skills \cite{akar2013bicycling, heinen2013effect}.  
Differences in preferences are typically stronger among occasional or non-cyclists than among regular cyclists \cite{aldred2017cycling}, thus suggesting that policies targeting women are particularly needed to increase cycling uptake. The main limitation of these studies is that they are mostly conducted via surveys or experiments with typically low sample sizes and/or a limited geographical breadth, and therefore low statistical explanatory power -- especially for the large number of possible confounders.

Recently, the emergence of new technologies for cycle-tracking and online-based services (e.g.~bike-sharing) generated an unprecedented stream of automatically collected data on cycling behavior, which enlarge the potential for research in this area. In this context, data from bike-sharing services have been used to study whether interventions to the bike-sharing facilities impacted men and women differently in the city of New York \cite{wang20191} and, more generally, to study factors affecting the demand for these types of services \cite{erenReview}. Data from Strava Metro, a service provided by the sport-tracking application Strava, have been used to study exposure to air pollution for different groups of cyclists in the city of Glasgow \cite{sun2017utilizing} and cycling patterns and trends for the city of Johannesburg \cite{musakwa2016mapping}. A few pioneering works have used GPS-based data to study route choices for different demographic groups in the city of San Francisco and Atlanta \cite{misra2018modeling, hood2011gps}.

In this study, we contribute to this strand of literature and use data from Strava to investigate the determinants of the gender gap in recreational cycling at a larger scale. With about 36 million users (2018 data) over 195 countries, Strava represents an unique data source on cycling-related behaviour \cite{StravaReport2018}, both in terms of the number of cyclists involved and the extent of the geographical coverage with a  methodologically homogeneous data collection. For this study, we collect and use data for over 60 cities in four geographical areas across the United States and Europe, to  explore the gender-cycling-gap at two different levels. First, we exploit the heterogeneity in the gender gap across the various cities in our dataset to challenge traditional hypotheses from the literature on the determinants of the gender gap in cycling. In particular, we study the strength of association between the gender gap in cycling measured at the level of urban centers and a set of urban indicators, spanning from morphological characteristics of the cities to safety indicators capturing the prevalence of cycleways and streets with low-speed limit in the road network. In the second part of the study, we move the analysis from a macro to a micro level. Here, focusing on the city of New York, we model the gender-cycling-gap measured at street-level in terms of specific urban features. By using logistic regression analysis, we investigate the association between the presence of dedicated cycling infrastructure and the volume of female cyclists on the street relative to men. The results indicate that streets with cycling infrastructures, particularly those ensuring the presence of physical separation for motorized traffic, are associated to a more balanced gender ratio, suggesting a way for policy makers to intervene to make urban environments more accessible for women. 

\section*{Results}
\subsection*{Using Strava data to measure the gender gap in recreational cycling}
We use Strava data to measure the gender gap in recreational cycling in 61 urban centers across four geographical areas: United States, United Kingdom, Italy and Benelux. 
Strava is an Internet service for tracking human exercise that relies on GPS data. The service supports up to 33 different activities, but it is mostly used for cycling and running. At the time of the data collection in 2018, Strava counted around 36 million users worldwide, corresponding to 0.6 billion recorded activities  \cite{StravaReport2018}. Of these, 284 millions were cycling-activities (47\%), and approximately one in five of cycling-uploads were by women (50 million)\cite{StravaReport2018}. Tracking of commuting is growing in popularity on Strava \cite{StravaReport2018}, however the majority of uploads refers to recreational and athletic cycling. The raw data consist of a collection of Strava segments, with information on users training on these from the associated leader boards. The data were processed to map gender and usage information from Strava segments to a network-based definition of streets and then aggregated for the entire city, following the pipeline described in the SI.

For each city $c$, we define the gender-cycling-gap as the ratio $\sigma_c$ between the total kilometers travelled by female cyclists and the overall kilometers travelled by cyclists of both genders. This measure accounts both for gaps in  trip-shares among men and women and for differences in travelled distances. By construction, $\sigma_c$ varies between 0 (no female cyclists) and 1 (no male cyclists): a value below 0.5 indicates the presence of a positive gender-cycling-gap (i.e. men cycling more than women). The closer the value to 0 the stronger the gap. For each geographical area covered by the study, Fig.~\ref{fig:citiesoverview} provides an overview of the cities included in the study, showing the five urban centers associated with the highest $\sigma_c$ for each area, as well as the location and the distribution of $\sigma_c$ of all covered cities, highlighting a persistent gender gap in recreational cycling (the full ranking is provided in the SI).  In our sample, the largest value for $\sigma_c$ is 0.21 in the municipality of Groningen, Netherlands, indicating the presence of a substantial gender gap in recreational cycling for all cities under consideration.   

Even within homogeneous geographical areas, we observe a substantial heterogeneity in $\sigma_c$ across cities. In the area of Benelux, in particular, $\sigma_c$ ranges between 0.06 (Charleroi, Belgium) and 0.21 (Groningen, Netherlands). Dutch cities (particularly those in the northern regions) generally outperform cities in
Belgium and Luxembourg. Among Italian cities, we observe a characteristic geographical pattern, with urban centers in the north-east displaying a lower gender gap than cities in the south and north-west. 
This north-south dichotomy is likely to be linked to the morphological characteristics of the country and the presence of a large flat land with a well-established cycling tradition. Differences in economic development might partially explain this structure as well.    
No geographical patterns are instead observable for cities in the United States and in the United Kingdom included in our sample. Interestingly, the is no evident link between the gender ratio and the size of a city. For instance, large cities such as Boston, Amsterdam and London perform high in the corresponding ranking, while top-ranking positions in Italy are dominated by relatively smaller urban areas.  
When comparing different countries, we stress that in general the penetration and typical use of Strava might differ across geographical areas. For instance, data on the use of Strava indicates different usage patterns and adoption rates in the United States compared to other countries \cite{StravaReport}. Therefore, we limit the comparison to cities within the same geographical area to ensure homogeneity in the Strava adoption by the general population, and we recommend to interpret a worldwide ranking with caution.  

\subsection*{Cross-city analysis of the gender gap}

\begin{figure}[t]
\centering
\includegraphics[width=1\linewidth]{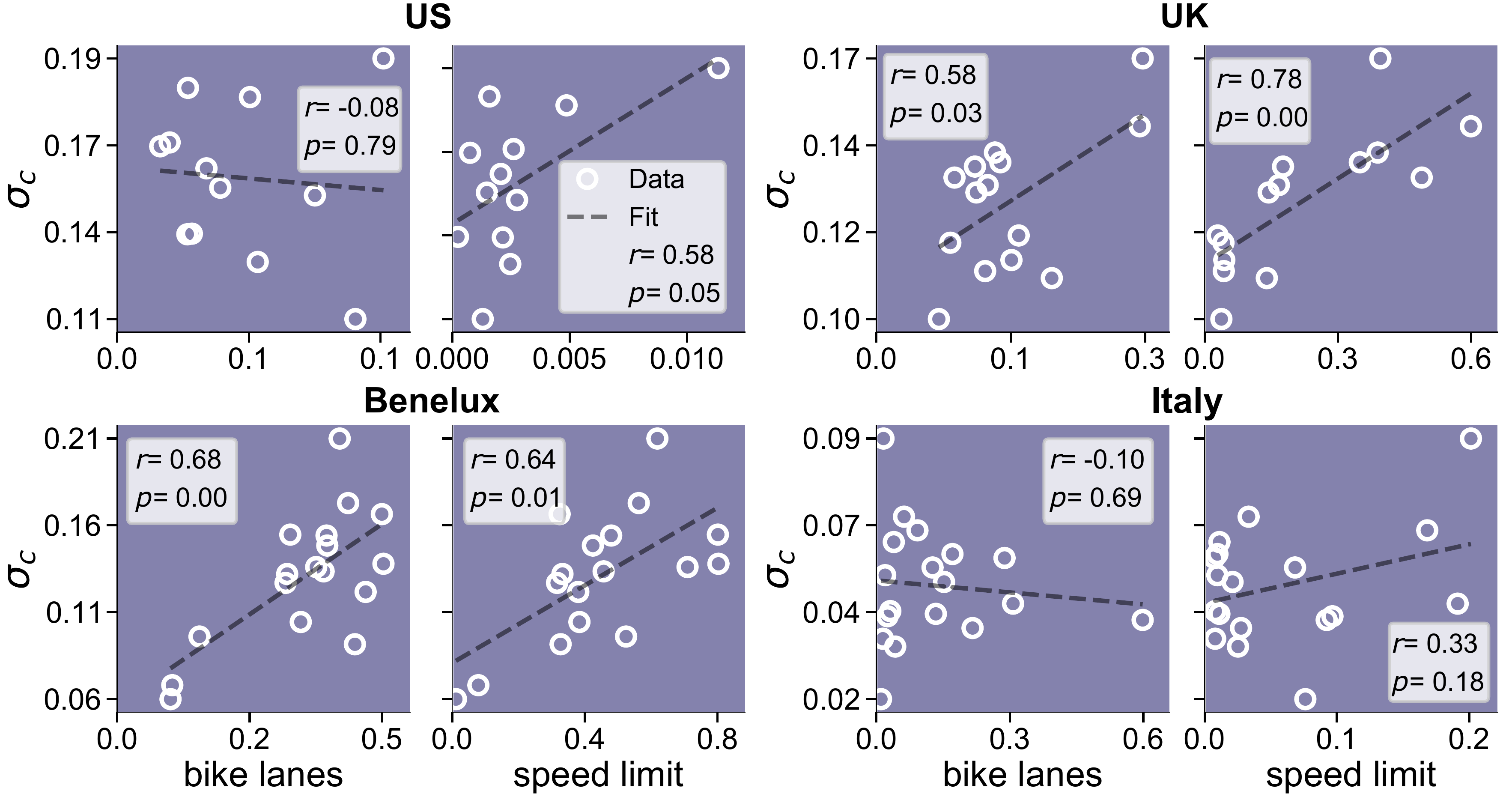}
\caption{\textbf{Correlations between gender ratio and urban road safety indicators.} The scatter plots show the correlations between two urban road safety indicators and the gender ratio $\sigma_c$, for cities in the four geographical areas separately. Each data point represents a city. The black line is the linear fit. The two urban road safety indicators capture the density of streets with a cycle lane in the street-network (indicator: $bike\;lanes$) and the density of streets with a speed limit up to 20 mi/h or 30 km/h in the street network (indicator: $speed\;limit$). A formal definition of the two indicators is provided in the SI.}
\label{fig:correlations}
\end{figure}

The survey-based literature on the gender gap in cycling suggests that women are more-risk averse, which would result in a lower cycling rate than men in environments perceived as risky \cite{aldred2017cycling}. Following this hypothesis, we investigate the association between the gender ratio $\sigma_c$ and two indicators of urban road safety, constructed using OpenStreetMap (OSM) data \cite{OpenStreetMap}. The first indicator (hereafter
 $bike\:lanes$) measures the proportion of streets with cycleways (either protected or unprotected) in the street network. The second metric (hereafter
 $speed\:limit$) provides the proportion of streets with a speed-limit equal or lower than $20\,\mathrm{mi}/\mathrm{h}$ or $30\,\mathrm{km}/\mathrm{h}$. Both metrics are weighted using the length of each street. 
Figure \ref{fig:correlations} reports the scatter plots between the gap $\sigma_c$ and the two urban road safety metrics, for the four main geographical areas separately. Each marker corresponds to a city, the black line is the linear fit. Both measures of road safety display a positive correlation with the observed gender ratio for the UK and for the area of Benelux. For cities in the United States, a positive (but weaker) correlation is only observable for the $speed\:limit$ indicator. For Italian cities, in contrast, both correlations are not statistically different from 0. %(at neither a significance level of 0.05 nor 0.1). 
This lack of significant correlations may be due to the presence of a large number of Italian cities with a very low degree of development of dedicated cycling infrastructure compared to cities in the UK and Benelux. Further, although some of these positive correlations appear to be driven by the presence of extreme values (Fig. S2 in the SI provides a similar picture without outliers), these are legitimate observations as the data were collected similarly for all cities.
Although limited to specific geographical areas, the positive correlations suggest an association between the degree of road safety and $\sigma_c$, thus supporting the hypothesis that low levels of women engagement with cycling may be explained by a greater concern for safety compared to men.

To untangle the effect of confounding factors, we explore the relationship between $\sigma_c$ and the two indicators of urban road safety controlling for a range of city-level indicators. To provide a thorough characterization of each city, the indicators are chosen from four domains: 1) \textit{E: Environment}, such as share of population in green areas, 2) \textit{BEI: Built-Environment \& Industrialization}, such as concentration of PM$\,$2.5, 3) \textit{SED: Socio-Economics \& Demographics}, such as GDP per person, and 4) \textit{M: Street Morphology}, such as average street grade. A full list of indicators is provided in Table~\ref{tab:City-level features}. The correlation matrix of the indicators across the entire sample is provided in Figure S3 in the SI. We include geographical dummies for the macro areas to account for different penetration levels of Strava worldwide.

\begin{figure}[t]
\centering
\includegraphics[width=1\linewidth]{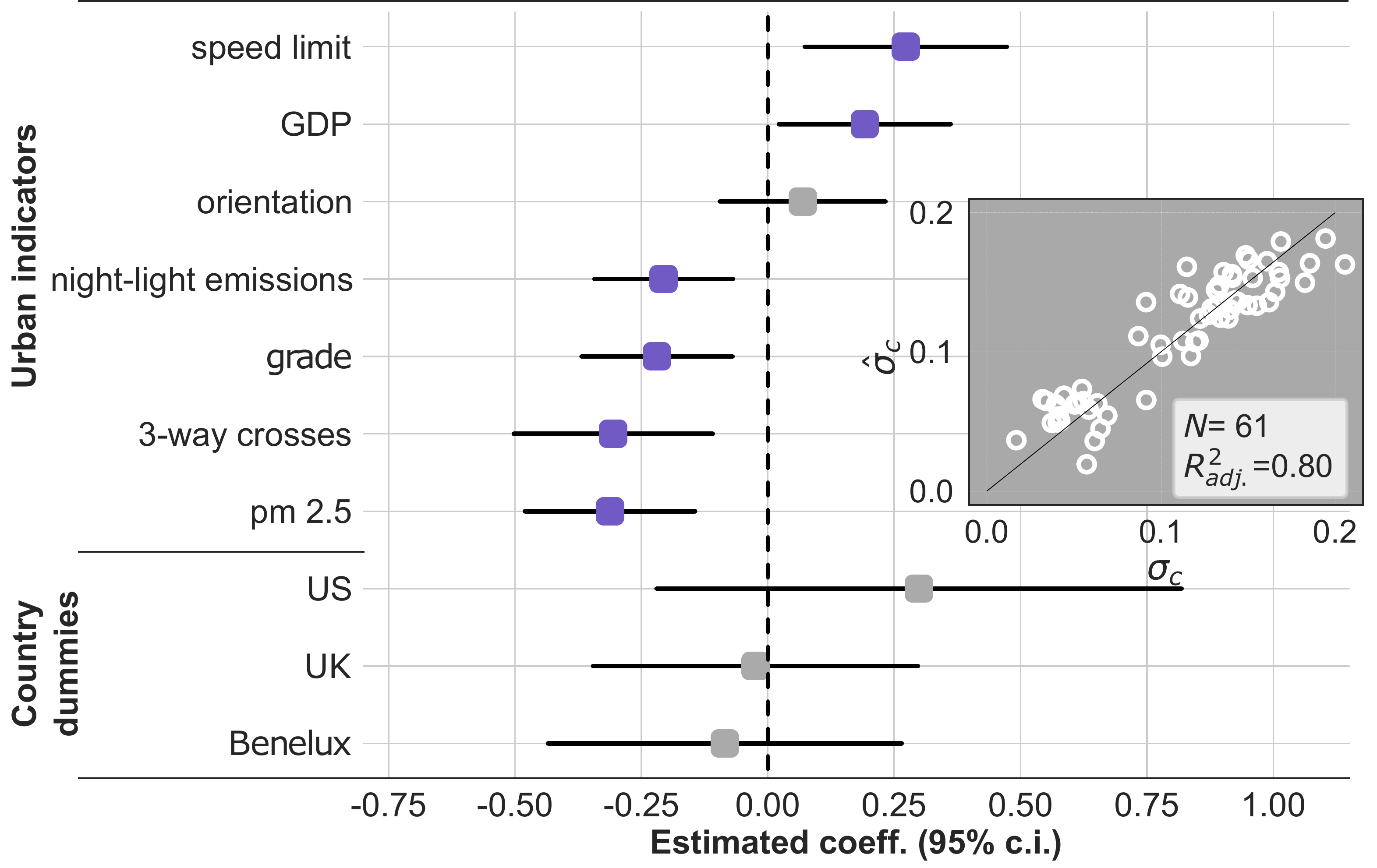}
\caption{\textbf{Results of regression analysis.} The main plot shows the estimated coefficients (square markers) with 95\% confidence intervals (black lines) for the final set of regressors included in the model. Model estimated via Ordinary Least Square, selection performed via exhaustive search. Selection criterion: Akaike Information Criterion (AIC). Statistically significant coefficient at 0.05 significance level are pictured in purple. The scatter plot displays the observed $\sigma_c$  vs the fitted $\sigma_c$. }
\label{fig:citylev_analysis}
\end{figure}

Coefficients (and 95\% confidence intervals) of a linear regression model estimated via Ordinary Least Squares (OLS) are shown in Fig. \ref{fig:citylev_analysis}, with statistically significant coefficients at 0.05 level (two-tailed test) pictured in purple. Information on the model selection is provided in the Materials and Methods. The regression analysis confirms the positive association between the gender ratio of cyclists and the $speed\;limit$ indicator. This association means that urban centers with a relatively wider low-speed zone typically present a more balance cycling uptake between men and women, after controlling for other confounding factors. Under the assumption that a wider low-speed zone indicates a less risky environment, this result also confirms that women are more susceptible than men to the perceived level of risk of the cycling environment.
Other insights emerge from the analysis of the control variables. First, we observe a negative association between $\sigma_c$ and the proportion of 3-way crosses. From a topological view-point, cities with a high proportion of 3-way intersections deviate from grid-like street networks, that, by contrast, present a large prevalence of (mostly orthogonal) 4-way intersections \cite{boeing2021street}. This result can be interpreted again under the lens of the degree of safety of the urban environment for cycling. Indeed, the literature has shown that not only are crashes involving cyclists more likely to happen at non-orthogonal crosses than at right intersections, but the former are more likely to lead to severe injuries \cite{asgarzadeh2017role}. 
Another key urban feature relates to the morphology of the street-network. The negative association between $\sigma_c$ and the $grade$ indicator shows that hillier cities display a larger gender gap in recreational cycling, controlling for all other factors. This result aligns with previous findings that women would have a preference for flatter routes \cite{hood2011gps} which may indicate a structural limit in the potential for cycling uptake by women in particular urban environments.
Interestingly, the analysis also indicates a lower gender ratio in cities with worse air quality (higher concentration of PM$\,$2.5). In absence of a (quasi-)experimental setting, however, we are unable to determine whether the air quality is a relevant feature per se or if it acts as a proxy for other city-level characteristics such as motorized traffic. Finally, the results indicate a more balanced cycling uptake between men and women in relatively wealthier cities (with a larger GDP per person) and cities with a lower degree of night-light emissions (which can be a proxy for the size of the city).
To test the robustness of this analysis, we estimate three additional models where we adopt different strategies to account for the different levels of penetration of Strava worldwide. These strategies differ in terms of: geographical coverage, specification of the geographical dummies and standardization of the input and target variables (Fig. S4 in SI). The results are largely consistent with the preferred specification provided here.

\subsection*{Street-level analysis of the gender gap}

\begin{figure}[t]
\centering
\includegraphics[width=\linewidth]{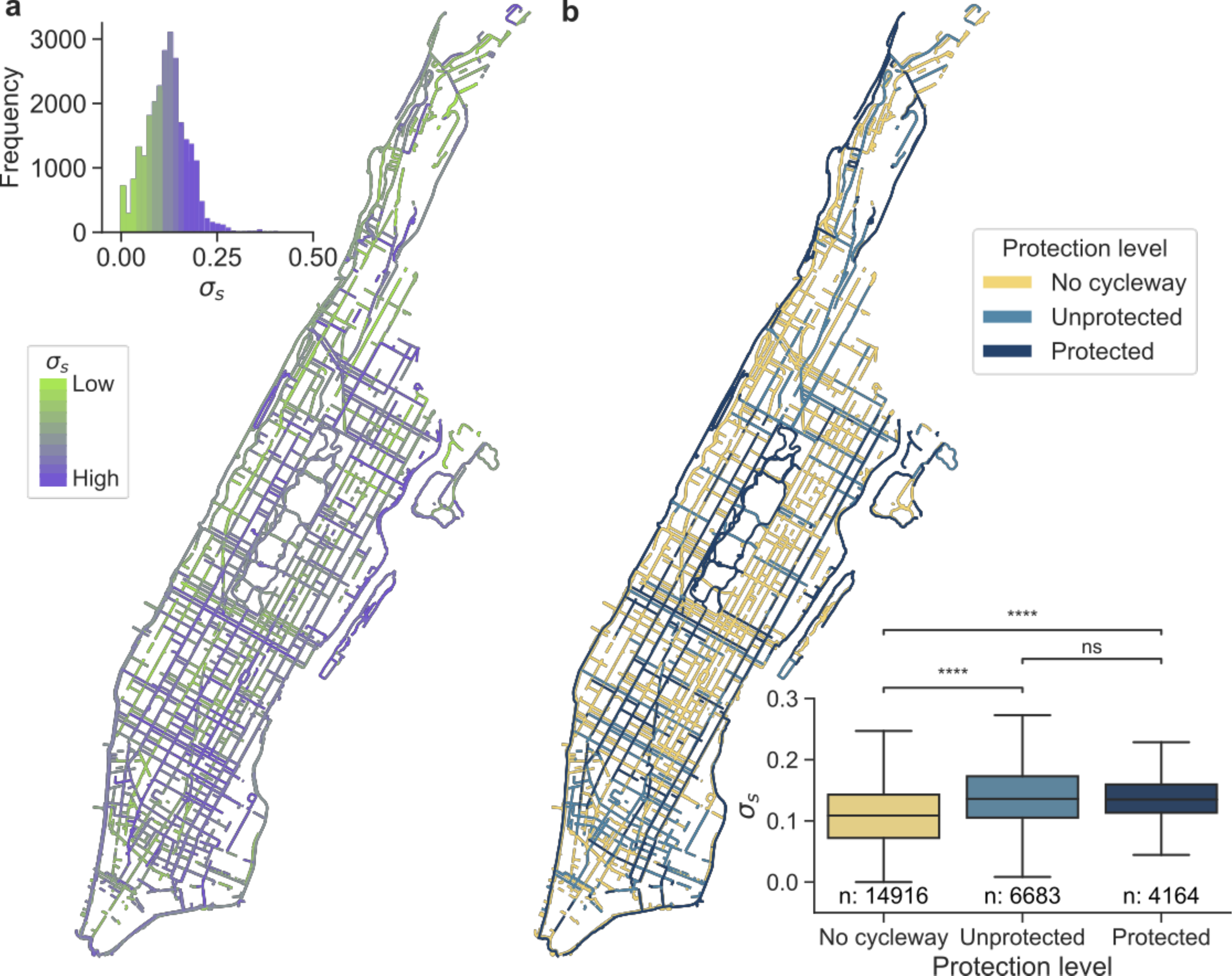}
\caption{\textbf{Streets characteristics and $\sigma_s$: The case of New York City.} a) The map displays streets in the borough of Manhattan included in the final sample. A 10-quantile color scheme has been used for the value of $\sigma_s$. The inset is the distribution of $\sigma_s$ for streets included in the final sample (computed over the entire city of New York).  b) The map displays the protection level of streets in the borough of Manhattan included in our sample. Yellow: no cycleway, light-blue: unprotected cycleway, dark blue: protected cycleway. The inset displays the box plots of $\sigma_s$ for streets with different levels of protection (computed over the entire city of New York). 'ns','*', '**', '***', '****' indicate the significance level of a Mann-Whitney-Wilcoxon test two-sided with Bonferroni correction, with the following p-values thresholds: 1e-4:"****", 1e-3: "***", 1e-2: "**", 0.05: "*", 1: "ns". }
\label{fig:NY_descriptive}
\end{figure}

The results in the previous section show that aggregated urban features model well the heterogeneity of the gender gap in cycling observed across different cities. They also confirm and provide quantitative support to traditional hypotheses from the literature, which are typically grounded on small-sample survey-based analyses. Though informative and affirmative, the previous analysis leaves open the question: Where exactly do women prefer to cycle? Also, which concrete interventions could policy makers implement to enhance cycling for women?

\begin{figure}
\centering
\includegraphics[width=\linewidth]{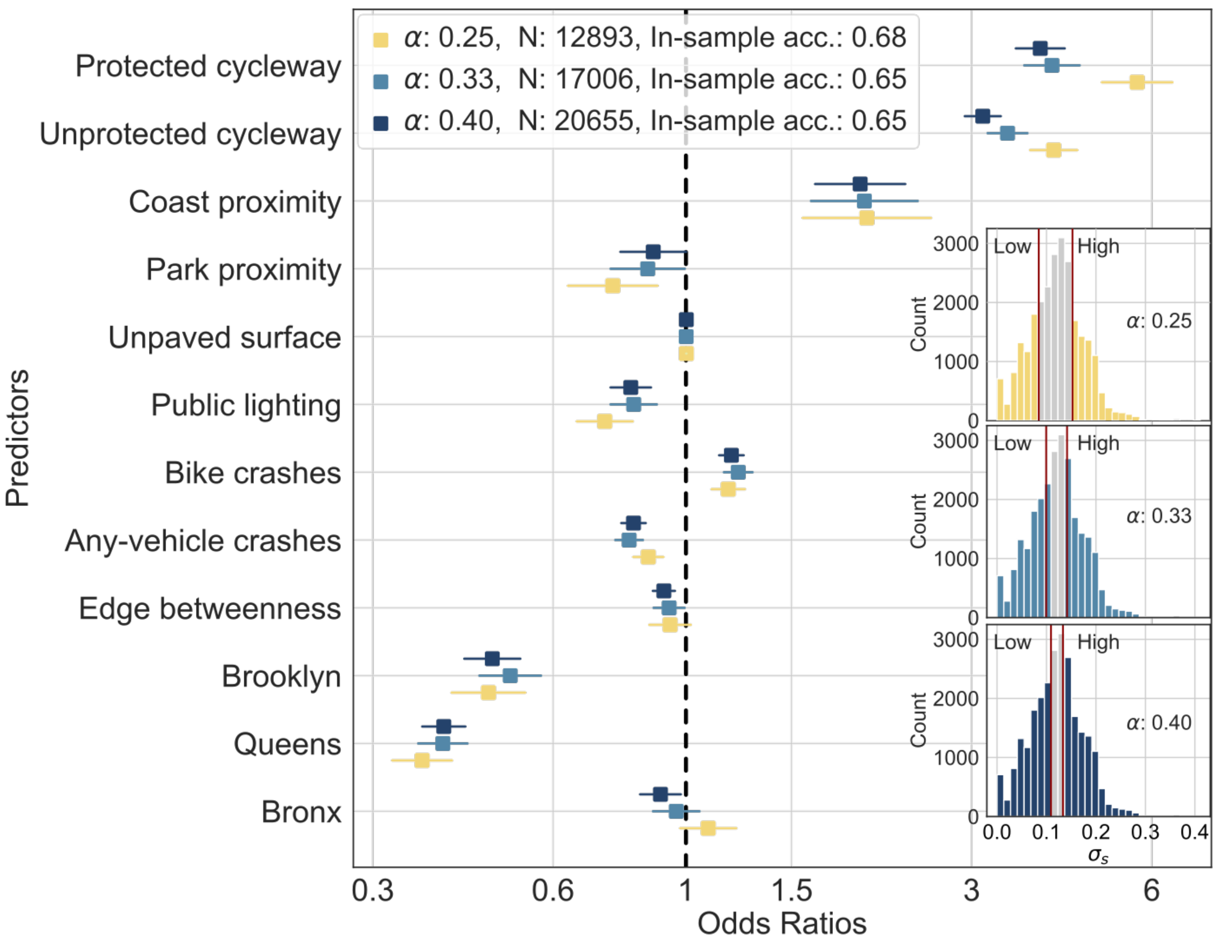}
\caption{\textbf{Odds Ratios of multivariate logistic regressions, for several level of the threshold $\alpha$} a) the chart presents estimated ORs for a multivariate logistic regression where the target variable is the binarized $\sigma_s$ and the predictors are listed in Table \ref{tab:Street-level features}. The squared dots are the point estimates. The straight lines are the estimated 95\% confidence interval for the corresponding OR. The model was estimated on three different sample selections, with the threshold $\alpha$ corresponding to 0.25 (yellow), 0.33 (light-blue) and 0.40 (dark blue). The ORs are computed exponentiating the corresponding estimated coefficients. For each estimated model, the legend reports the value of the threshold $\alpha$, the number of observations and the in-sample accuracy. b) The histograms show the mapping between the $\sigma_s$ and the binarized $\sigma_s$ for the three values of the threshold $\alpha$: 0.2 5 (yellow), 0.33 (light blue) and 0.40 (dark blue).}
\label{fig:NY_Logit}
\end{figure}

To answer these questions, we shift the focus from a macro-level comparison across cities, to a micro-level setting where the unit of analysis are streets within one city as opposed to the entire city itself. This shift in perspective allows us to examine the preferences of women for street-level characteristics in greater detail, thus identifying potential targets for interventions by policy makers. Among the available cities, we select as a case study the city of New York, whose large collection of administrative datasets represents an opportunity to enrich the analysis with data not otherwise available from OSM only. In particular, using OSM data, we are able to characterize each street in our sample with information on: the presence (or absence) of a protected (or unprotected) cycleway, the presence of public lighting, the type of surface (paved vs unpaved), whether the streets are close to a park or to a coastline. The administrative data are instead used to measure the number of crashes (any type of vehicles or bicycle-related only) on the street (normalized by the street length) and to associate each street to a neighborhood. Finally, to proxy for traffic flow, we compute the normalized edge betweenness \cite{latora2017complex} of each street in the street network. The edge betweenness is a network centrality measure capturing the number of the shortest paths that go through an edge in the network. A summary of all features is provided in Table \ref{tab:Street-level features}. As for the city-level analysis, we use Strava data on cycling to quantify female preferences for a street $s$. We measure the proportion of female cyclists out of all cyclists travelling via street $s$, and call this metric $\sigma_{s}$. The indicator $\sigma_s$ is a direct street-level extension of $\sigma_c$ -- indeed $\sigma_c$ can be constructed averaging over $\sigma_s$ with weights equal to the product between the length of each street and the total number of cyclists on it. The larger $\sigma_s$ the greater are womens' preferences to cycle on street $s$. Compared to a simple count of female cyclists, this relative measure has the advantage of quantifying female-specific preferences towards a street $s$, irrespective of the total level of \textit{popularity} of the street. Therefore, the metrics will not be distorted towards streets that are very popular for cyclists in general (for instance for their position in the street network), but that may not present features that are particularly appreciated by our target group. 
In addition, we adopt a data-driven approach to filter streets with a low number of cyclists (described in the SI). This filtering ensures that the observed $\sigma_s$ is computed on a sufficiently large cyclist base.
The distribution of $\sigma_s$ is bell-shaped with a mean around 0.12 and a range between 0.00 and 0.41 (\figurename \ref{fig:NY_descriptive}(a)). Stratifying the distribution by protection level of the street ('No cycleway', 'Unprotected cycleway', 'Protected cycleway') -Fig. \ref{fig:NY_descriptive}(d)- we see that streets with no forms of dedicated-infrastructure are typically associated with lower $\sigma_{s}$ than streets with either protected or unprotected cycleway: the median value of $\sigma_{s}$ for streets with no cycleway roughly corresponding to the 25th percentile of both the distributions of streets with protected or unprotected cycleways. This descriptive evidence provides a first indication that streets with some form of  cycling infrastructure are typically used more intensively by women than streets with no dedicated infrastructure at all. 
To delve deeper into women's preferences for dedicated cycling infrastructure, we study the degree of association between the presence of protected and unprotected cycleways and $\sigma_s$ by means of a logistic regression analysis. We classify streets into two classes, \textit{Low} and \textit{High}, corresponding to the bottom, and top 33\% of the distribution of $\sigma_{s}$ and estimate the Odds Ratios (OR) via multivariate logistic regression. To check the robustness of the results, the analysis is repeated choosing different thresholds $\alpha$ (0.25 and 0.40, instead of 0.33) for the classification. The results (presented in Figure \ref{fig:NY_Logit}) are consistent across sample specifications, with generally slightly larger estimates on more extreme samples (lower values of the threshold $\alpha$).   

The main result pertains to the role of dedicated cycling infrastructure. With an estimated OR of around 4.08 (95\% confidence interval: [3.67, 4.54]), the analysis indicates that the odds to be classified \textit{High} are more than four times greater for protected cycleways than for streets with no cycleway (used as baseline). This result is largely in line with the survey-based literature on the gender-cycling-gap, according to which women would favor physical separation more than men \cite{aldred2017cycling, garrard2008promoting, misra2018modeling}.
 Though smaller in magnitude, we estimate a similarly positive association between the presence of an unprotected cycleway and $\sigma_s$. This analysis suggests that, whenever  protected cycleways are not feasible due to either budget or physical constraints, the use of shared unprotected cycleways would still be a way to make the urban environment more accessible for female cyclists. In light of recent findings \cite{PEARSON2022101290, MARSHALL2019100539} which suggest that unprotected cycleways would not enhance the degree of safety of the road-network for cyclists, our results suggest that subjective safety may matter more than objective safety. 
 In terms of other control variables, in line with the assumption that women favor more quiet streets, we estimate an OR below 1 for our proxy for traffic-flow (\textit{Edge-betweenness}) and for the volume of accidents (by any type of vehicle). The positive association with the volume of bicycle crashes, on the other hand, is likely to be the effect of reverse causality: a more balanced gender ratio is typically associated with a larger volume of cyclists, with an increased likelihood of bicycle crashes. The two dummies on coast and park proximity, here inserted as a proxy for the natural environment within which the street is located, appear to have opposite effects, with an estimated OR above 1 for $Coast\;proximity$ and below 1 for $Park\;proximity$ (with the latter being only statistically significant at 0.05 level for $\alpha=0.25$). On one hand, the reason for the high coast proximity value is due to the morphology of the city of New York and the presence of a long protected cycleway along the coastline of Manhattan acting as an attractive infrastructure and impacting nearby streets too, with many cyclists riding through to reach it. On the other hand, the negative association with the $Park\;proximity$ dummy can be traced back to the location of the green areas under consideration, often in non-central locations (note that streets within the Central Parks largely fall into the excluded part of the distribution around the median value). Information on the presence of public lighting is generally very sparse in OSM and particularly for New York City (we assume public lighting to be absent only whenever explicitly stated, with less than 100 streets classified as without public lighting), therefore the negative estimated OR requires further analysis with more complete data. Finally, although hard to generalize to other urban contexts, we observe strong negative neighbourhood effects, particularly for the boroughs of Brooklyn and Queens (compared to the baseline borough of Manhattan). 
 
 The multivariate logistic regression presented in this paragraph allowed us to investigate the average impact of specific street-level features on the probability that a street belongs to the $High$ or $Low$ $\sigma_s$ group. The overall in-sample accuracy of the model oscillates between 0.68 and 0.65, depending on the threshold $\alpha$. To check the robustness of our analysis, we additionally compared the results obtained using the multivariate logistic regression to the results of a Random Forest classifier. The forest-based classifier provides a higher accuracy (average out-of-sample accuracy of 0.83 for $\alpha$ = 0.33 compared to 0.65), at the expense of a lower degree of explainability. Nevertheless, using the game-theory inspired concept of shapely values \cite{lundberg2017unified}, for each street, we can quantify the impact on the prediction of the considered features. In Fig. S6 in the SI,  we present shapely values computed on a random selection of 500 data points, which largely confirm the results of the logistic regression in terms of the central role of the dedicated cycling infrastructure.

\section*{Discussion}
In this study, we investigated the determinants of the gender-cycling-gap using data for over 60 cities in Europe and the United States. Unlike the vast majority of previous analyses that used survey-based data, we leveraged large automatically collected data from the online sport-tracking application Strava. We first related female cycling rates in different European and American cities to city-level characteristics and found evidence for traditional hypotheses which link the observed gender gap in cycling to gender specific preferences on road safety. Additionally, we found higher female cycling rates in flatter than in hillier cities, also in line with the literature \cite{hood2011gps}. This is an interesting result as there may be structural, morphological or cultural \cite{goddard2014gda} constraints for specific places where the cycling uptake is harder to increase for women. For urban planning this result suggests that ad-hoc infrastructural interventions such as the provision of cycleways or the enlargement of the low-speed limit zones could have limited efficacy in these contexts and may require concurrent behavioral incentives, for instance to expand the adoption of e-bikes. A novel result concerns the strong association between the gender-cycling-gap and the air quality of a city, which however requires further research within a (quasi-)experimental setting.

In the second part of the study, we shifted the focus from a macro comparison across cities to a micro-level analysis, at the level of single streets. If the first analysis successfully provided evidence for and expanded existing hypotheses (further validating our data as a reliable source on cycling behavior), the second aims at capturing the role of urban features modelled at a higher resolution and delving deeper into the association between the gender-cycling-gap and the presence of dedicated cycling infrastructure. We selected the city of New York as case study for this component of the study. Using multivariate logistic regression analysis, we have shown the existence of a positive association between the volume of female cyclists (relative to men) and the presence of dedicated cycling infrastructure. 
The positive association between $\sigma_s$ and the presence of a protected cycleway was expected and well-documented in the literature, which highlights the strong preference of women for physical separation from motorized traffic \cite{aldred2017cycling, garrard2008promoting, misra2018modeling}. More novel and interesting is the observed association with the presence of an unprotected cycleway. In light of recent studies showing that unprotected cycleways may not enhance the degree of objective road safety \cite{MARSHALL2019100539, PEARSON2022101290}, our result suggests that the perceived degree of safety may induce women to cycle more than the actual degree of safety. 
Therefore, in contexts where no physical separation is possible (for instance for space or budget constraints), the provision of shared cycleway may still act as a way to make to urban environment perceived as more accessible by women. However, given that the increase in the perception of safety induced by this type of infrastructure may not always translate into a lower risk cycling environment, the planning of this type of infrastructure should be evaluated carefully by city planners, for instance favoring specific solutions associated to greater safety levels. 

Overall our study validated survey-based results quantitatively using unprecedentedly large-scale automatically collected data. With around 36 million users worldwide in 2018, Strava was among the major applications for sport tracking and as such, a reliable information on cycling-behavior for regular cyclists. The main limitation of our study pertains to the representativeness of Strava users and the purposes of Strava trips. For example, having a considerable gender gap in the Netherlands (Fig.~\ref{fig:citiesoverview}), contrary to expectations \cite{pucher2008making}, the Strava data are clearly not representative, and neither users nor purposes of use can be clearly inferred. We therefore stress that Strava does not necessarily reflect recreational cycling only, and that such assumptions should be challenged and explored with richer data sets or qualitative methods. Nevertheless, we did our best to account for the representative challenge of this data set, first by comparing only cities in the same region, and second by comparing streets only in the same city, aiming to minimize user and trip purpose variation. A second limitation of the Strava data set is the inability to extract the potentially useful information of cyclist volumes \cite{aldred2016does}, as the raw data are not individual cycling traces but Strava segments with only aggregated statistics. This aggregation also implies that the same cyclists may cycle on many segments in one or multiple sessions and we would not be able to identify them.

It is unclear to which extent our results can be generalized to cycling for purposes other than recreational, such as transport, and to less-skilled cyclists (occasional cyclists and not cyclists). It is therefore important to find data sources that are able to reliably distinguish between such purposes and users, since gender-based constraints can differ between these categories \cite{heesch2012gdr}. However, since the survey-based academic literature on gender-cycling-gap indicates that cycling preferences differ less among regular cyclists than among occasional ones \cite{aldred2017cycling,prati2019gender}, the results of our analysis could be interpreted as a lower-bound and it is likely that the identified factors play an even larger role in explaining the gender-cycling-gap in the general population. Another limitation concerns the cross-sectional nature of the available cycling data. The absence of a longitudinal dimension limited the extent to which temporal variations could be analysed in the data, hindering the use of policy-evaluation statistical tools such as diff-in-diff techniques to evaluate casual effects along with correlations. 

Finally, there is a variety of gender-specific constraints apart from street safety that future studies should take into account, from cultural and psychological reasons \cite{aldred2016does,goddard2014gda}, to other environmental factors and harassment by motorists \cite{heesch2012gdr,graystone2022gendered}. Gender inequality and gendered transport habits may also play a large role, such as more frequent trip chaining by women due to childcare and other errands \cite{prati2018gender,garrard2012women}. Therefore, while street safety and urban design are undoubtedly important ingredients, there is no universal, simple fix for getting rid of the gender gap in cycling towards more sustainable mobility. It remains a complex societal issue that needs to be tackled from multiple angles \cite{hanson2010gmn}.

\section*{Materials and Methods}
\paragraph{Data sources}
The study used data from multiple sources summarised below. More detailed information on the collection, cleansing and processing of each of these strands of data is provided in the SI.
\begin{itemize}
    \item Strava data on biking extracted from Strava heat maps using the public API in November 2018.
\end{itemize}
\textbf{Cross-cities analysis}
\begin{itemize}
 \item City-indicators from the Global Human Settlement - Urban Centre Database 2015 (GHS-UCDB R2019A) \cite{GHSUrbanCentreDatabase}. 
    \item Street-network indicators for the urban centers of interests extracted from \cite{boeing2021street}.
    \item OpenStreetMap \cite{OpenStreetMap} data extracted though the Python library OSMnx \cite{boeing2017osmnx} to compute urban safety indicators measuring: 1) the proportion of the street network with bike-lanes, and 2) the proportion of streets with low speed limit. Detailed information on the extraction pipeline is provided in the SI.

\end{itemize}
\textbf{Case study on the City of New York}
\begin{itemize}
    \item OpenStreetMap \cite{OpenStreetMap} data on street-level characteristics extracted during the process of remapping of Strava data via the python library OSMnx \cite{boeing2017osmnx}. For each street, we retained information on: the presence of public lighting, the presence of protected or unprotected bike-lanes, proximity with a park or with the coastline and whether the surface is paved. A full list of the OSM tags used is provided in Table S2.
    In addition, for streets in the largest component of the street network, we computed the edge-betweenness via the python library $graph-tool$. Streets outside the largest component of the network (i.e. excluding streets in the borough of Staten Island) were excluded from the sample.
    \item Administrative data from the OpenData Portal of the city of New York on location of all (any-vehicle) accidents and bike accidents \cite{NYCadmin}.
    \item Shapefiles of the administrative boundaries of boroughs in the city of New York. Available at \cite{NYCadmin}.
\end{itemize}

\paragraph{Ordinary Least Squares regression}
We estimate a linear regression model via Ordinary Least Squares (OLS) of the form: 
\begin{equation}
    \sigma_c=\sum_{j=1}^{N}\beta_{j}z_{j,c}+\epsilon_{c} \quad  c=1,..,61
\end{equation}
where the list of regressors $z_{j}$ includes: \textit{speed limit}, \textit{orientation}, \textit{GDP},  \textit{3-way crosses}, \textit{night-light emissions}, \textit{grade}, \textit{pm2.5} plus three dummy variables for the macro area to which the city belong (US, UK, Benelux, \textit{baseline}: Italy).  Prior to undertake any analysis, we perform a z-score-transformation of all regressors. As such, the results should be interpreted in terms of standard deviations. Out of the initial 15 city-level indicators collected (provided in Table 1) , the final subset of seven indicators (plus the three country-level dummies) included in the regression were selected via exhaustive search to minimize the Akaike Information Criterion (AIC) of the model. For estimation of the Linear regression, we use the OLS function of the Python library $statsmodel$ \cite{Seabold2010StatsmodelsEA}.

\paragraph{Multivariate logistic regression}
To assess the degree of association between $\sigma_s$ and the presence of cycling dedicated infrastructure, we use a multivariate logistic regression. We restrict the sample to streets belonging to the bottom and top 33\% of the distribution of $\sigma_s$ and classify streets in $Low$ and $High$ $\sigma_s$, respectively. As a robustness check, the analysis is repeated for alternative values of this threshold (25\% and 40\%). We use features described in Table \ref{tab:Street-level features} as predictors and the binarized $\sigma_s$ as the target variable. Moreover, we
scale continuous predictors (\textit{Any-vehicle crashes}, \textit{Bike crashes} and \textit{Edge-betweenness}) using a z-score-transformation to normalize the magnitude of the estimated coefficients.
For estimation of the Multivariate Logistic regression, we use the Logit function of the Python library $statsmodel$ \cite{Seabold2010StatsmodelsEA}. The OR for each predictor is computed exponentiating the corresponding estimated coefficient. 

\paragraph{Classification task: models comparison}
We evaluate the accuracy of the multivariate Logistic Regression and the Random Forest classifier using a 80-20 train-validation split of the sample selected using $\alpha=0.33$. Hyper-parameters setting via grid-search optimization is performed for the Random Forest classifier ( Selected parameters: $'max\_features'$: $'auto'$,$'min\_samples\_leaf'$: 1,$'min\_samples\_split'$: 2,$'n\_estimators'$: 150). A stratified 10-fold approach is then adopted to evaluate the variability in the accuracy of the two models. For this strand of the analysis, we used the scikit-learn library \cite{scikit-learn} for Python. 

\section*{Code availability}
The Python code developed for the data analysis is available at: \url{https://github.com/alibatti/GenderCyclingGapUsingStrava}.

\section*{Acknowledgement} We thank Ane Rahbek Vierø for helpful discussions. PB, APa, APe acknowledge partial support from Intesa Sanpaolo Innovation Center. The funders had no role in study design, data collection and analysis, decision to publish, or preparation of the manuscript.

%%%%%%%%%%%%%%%%%%%%%%%%%%%%%%%%%%%%%%%%%%%%%%%%%%%%%%%%%%%%%%%%
% BIBLIOGRAPHY
%%%%%%%%%%%%%%%%%%%%%%%%%%%%%%%%%%%%%%%%%%%%%%%%%%%%%%%%%%%%%%%%
%\bibliography{biblio}


\clearpage

\begin{turnpage}

\begin{table}
	\centering
	\caption{Description and source of indicators at the level of urban center included in the study, by category. \footnote{Categories: E: Environment, BEI: Built-Environment and Industrialization, SED: Socio-Economics and Demographics, M: Street Network Morphology, RS: Road Safety. *Indicates that the data from the original data sources required specific processing described in the SI.}}
%		\resizebox{\linewidth}{!}{%
		\begin{tabular}{p{0.1\linewidth} | p{0.1\linewidth} | p{0.7\linewidth} | p{0.1\linewidth}}
			\toprule
			\textbf{Category} &
			\textbf{Variable name}  &
			\textbf{Description} &
			\textbf{Data source} \\  
			& $\sigma_c$ & Proportion of kilometers rode by female cyclists to the overall kilometers rode by any cyclist within the urban area & Strava* \\
			 
			E & share green & Share of population living in the high green area in 2015 in the
			Urban Centre of 2015. Ranging between 0-1 & \cite{GHSUrbanCentreDatabase} \\
			& open space & Percentage of open-spaces within the spatial domain of the  Urban Centre. Ranging between 0-100 & \cite{GHSUrbanCentreDatabase}  \\
			 
			BEI & built area &  Amount of the built-up area per person in 2015 calculated within the
			spatial domain of the Urban Centre. Expressed in square meters per
			person & \cite{GHSUrbanCentreDatabase}  \\
			& light emissions & Average night time night-light emission calculated within the Urban Centre
			spatial domain. Expressed in nano-watt per steradian per square centimetre & \cite{GHSUrbanCentreDatabase}  \\
			& pm2.5 &  Total concentration of PM2.5 for reference epoch 2014, calculated
			over the Urban Centre. Expressed in $\mu g/m3$ & \cite{GHSUrbanCentreDatabase}  \\
			 
			SED & area & Area of the spatial domain of the Urban Centre. Expressed in square meters & \cite{GHSUrbanCentreDatabase}  \\
			& population & Population density within the spatial domain of the Urban Centre & \cite{GHSUrbanCentreDatabase} * \\
			& GDP & GDP per capita for year 2015 within the Urban Centre. Expressed in US dollars & \cite{GHSUrbanCentreDatabase} * \\
			 
			M & degree & Average node degree of street network within the spatial domain of the Urban Centre & \cite{boeing2021street} \\
			& grade & Average absolute inclination of streets within the spatial domain of the Urban Centre. Expressed in percentage & \cite{boeing2021street}  \\
			& orientation & Orientation order of street network bearings within the spatial domain of the Urban Centre. & \cite{boeing2021street}  \\
			& 3-way crosses & Proportion of nodes that represent a 3-ways street intersection in the street network within the spatial domain of the Urban Area. Ranging between 0-1 & \cite{boeing2021street}  \\
			& straightness & Ratio of straightline distances to street lengths for streets in the street network within the spatial domain of the Urban Area & \cite{boeing2021street}  \\
			 
			RS & bike lanes & Proportion of streets with cycleways (either protected or unprotected) computed on streets within the spatial domain of Urban Centre & OSM* \\
			& speed limit & Proportion of streets with a speed-limit equal or lower than 20 $mi/h$ or $30 km/h$ computed on streets within the spatial domain of Urban Centre & OSM*  \\
			\toprule
		\end{tabular}
	\label{tab:City-level features}
\end{table}

\clearpage

\begin{table}
	\centering
	\caption{List of street-level urban features included in the New York City case study.\footnote{\textsuperscript{1} Information on the presence of public lighting is very sparse in OSM. In case of missing information, we assumed that the public lighting is available. *Indicates that the data from the original data sources required specific processing (e.g. for normalization) described in the SI.}}
%		\resizebox{\linewidth}{!}{%
		\begin{tabular}{p{0.15\linewidth} | p{0.7\linewidth} |p{0.15\linewidth}}
			\toprule
			\textbf{Variable name} & \textbf{Description} &
			\textbf{Data source} \\  
			Unprotected cycleway & Dummy for presence of a shared or unprotected bike-lane &
			OSM \\
			Protected cycleway & Dummy for either the presence of a protected bike-lane or streets with no vehicles &
			OSM \\
			Public lighting\textsuperscript{1} &Dummy for the presence of public lighting &
			OSM \\
			Unpaved surface &Dummy for unpaved surface &
			OSM  \\
			Park proximity & Dummy for streets next to a park (within 15 meters) & 
			OSM*  \\
			Any-vehicle crashes & Number of crashes involving any type of vehicles per 10m of street length &
			NYC*  \\
			Bike crashes & Number of bicycle crashes per 10m of street length &
			NYC*\\
			\{borough\_name\} & Dummy for boroughs (baseline: Manhattan)&
			NYC  \\
			Coast proximity & Dummy for street next to the river coast&
			OSM* \\
			Edge betweenness & Edge betweenness of the streets computed for streets within the largest component of the street network & \\
			\toprule
		\end{tabular}
	\label{tab:Street-level features}
\end{table}
	
\end{turnpage}
%%%%%%%%%%%%%%%%%%%%%%%%%%%%%%%%%%%%%%%%%%%%%%%%%%%%%%%%%%%%%%%%

\newpage

%%%%%%%%%%%%%%%%%%%%%%%%%%%%%%%%%%%%%%%%%%%%%%%%%%%%%%%%%%%%%%%%
% SUPPLEMENTARY MATERIAL
%%%%%%%%%%%%%%%%%%%%%%%%%%%%%%%%%%%%%%%%%%%%%%%%%%%%%%%%%%%%%%%%

\clearpage
%%%%%%%%%% Prefix a "S" to all equations, figures, equaitons, tables and reset the counter %%%%%%%%%%
\setcounter{figure}{0}
\setcounter{table}{0}
\setcounter{equation}{0}
\makeatletter
\renewcommand{\thefigure}{S\arabic{figure}}
\renewcommand{\theequation}{S\arabic{equation}}
\renewcommand{\thetable}{S\arabic{table}}
%%%%%%%%%% Prefix a "S" to all equations, figures, equations, tables and reset the counter %%%%%%%%%%

\setcounter{secnumdepth}{2} %%%%%%%%%% ADD NUMBERED SECTIONS - PRL DOESN'T WANT IT!

\widetext
\begin{center}
	\textbf{\large Supplemental Material:\\ Revealing the determinants of gender inequality in urban cycling with large-scale data}
\end{center}
\section{Strava data on recreational cycling: data collection and processing} \label{sec:StravaData}
\subsection{Data collection}
Raw Strava data consist of a collection of Strava segments for 62 cities located in four geographical areas: the United States, United Kingdom, Benelux (Belgium, Netherlands and Luxembourg) and Italy. For the sensitivity analysis only, the dataset was extended to include 8 additional cities across other European countries. A Strava segment is a single portion of a road or a trail upon which Strava users compete by recording their times. The performance of a user exercising upon a segment is automatically recorded into its leader board, which in turn provides a picture of the characteristics of users exercising on a specific trail. Each raw data record consists of geographic information in the form of a linestring of lat-long coordinates plus two statistics extracted from the associated leaderboard:
\begin{enumerate}
    \item the total number of unique cyclists training on the segment. This information corresponds to the sum of the length of the female and male leader boards, (it should be noted that -irrespective of the number of training performed on the segment- each cyclist is only included once in the corresponding leader board, according to their best performance on the segment);
    \item the gender split of users training on the segment, in terms of the length of the female and male leader boards respectively (corresponding to the number of unique female and male cyclists training on the segment).
\end{enumerate}.

The data collection comprised of two phases, both undertaken in November 2018. In the first phase, we collected the whole corpus of segments (\~ 16.4 million) available at the time through the Strava API. This step provided us with a series of information related to each segment, in particular its unique ID and the linestring geometry. The second phase, consisted in the collection of data from the female and male leader boards (with two separate queries for the same ID) associated with each segment. In particular, given a city, we made queries for the leader boards of all the segments whose geometry is contained for at least 75\% within the city boundary (extracted from OpenStreetMap). The information from the leader board was then processed to extract the statistics described above.

\subsection{Characteristics of raw Strava segments}
\begin{figure}
\includegraphics[width=\textwidth]{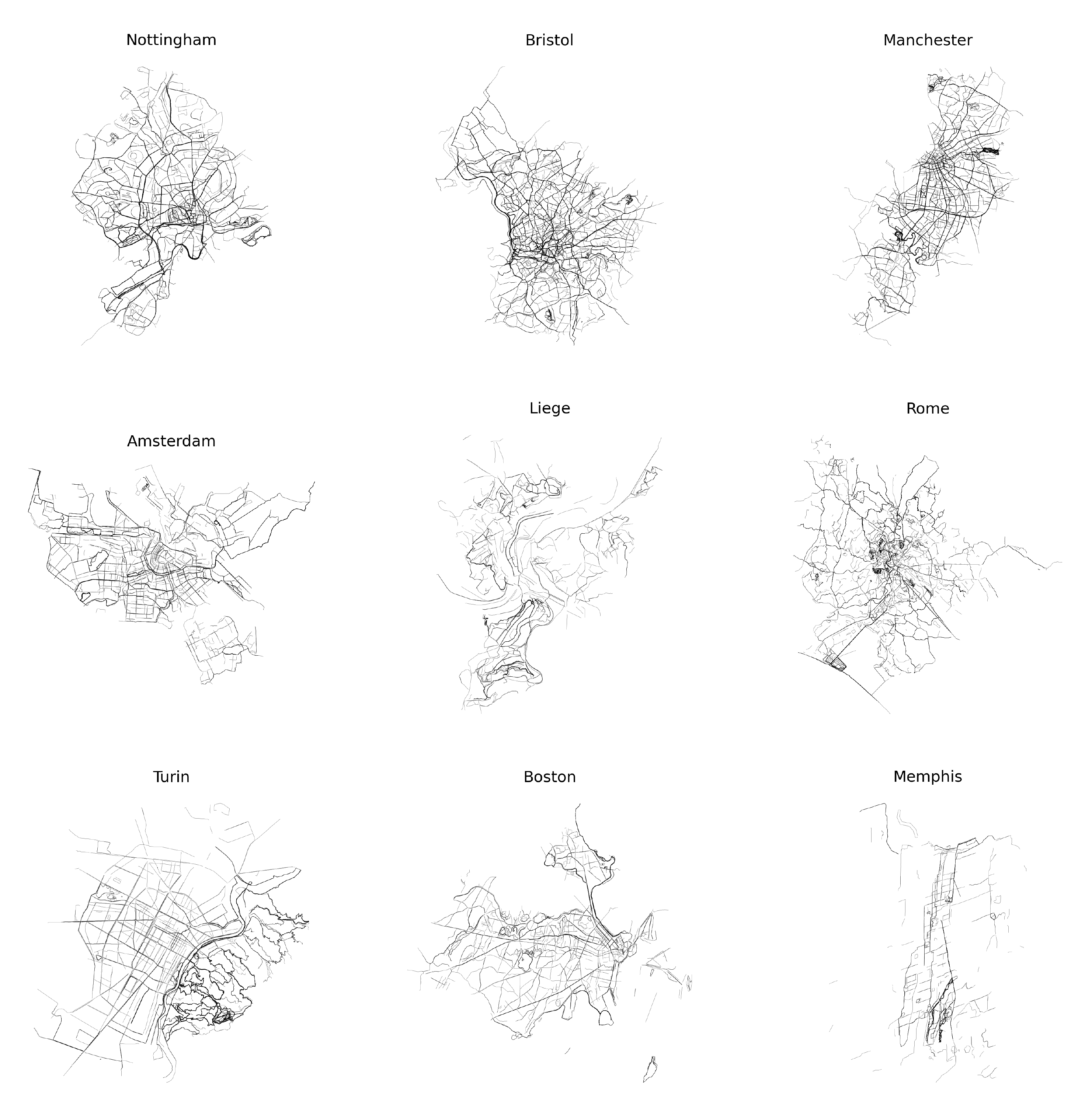}
\caption{\textbf{Visualisation of raw Strava segments for nine cities.} Graphical visualisation of raw Strava segments for nine cities for the four main geographical areas covered by the study (United Kingdom: Nottingham, Bristol and Manchester; Benelux: Amsterdam and Liege, Italy: Rome and Turin, United States of America: Boston and Memphis). All segments are plotted with the same color intensity. Darker lines indicate the presence of overlapping segments.}
\label{fig:cities_raw_strava_segments}
\end{figure}

Strava segments are not pre-defined by the app developers but they are directly generated by the users of the app. The result of this user-driven generating process is a set of segments highly heterogeneous in length, both across cities and within the same city. Indeed, some segments may correspond to portions of a street, while others define long trails spanning multiple streets. Furthermore, segments can partially or completely overlap each others. An illustration of the extent of this characteristic is provided in \figurename~\ref{fig:cities_raw_strava_segments}, where we present a visualisation of raw Strava segments for nine cities. As segments are plotted with the same color intensity, darker areas on the map indicate a series of overlapping segments. Furthermore, 
\tablename~\ref{tab:Strava_raw_descr_a} provides summary information on the raw data for the cities included in the final sample for the study. In terms of length of segments, the table depicts well the discussed heterogeneity both across cities and within the same urban center. An illustration of the large heterogenity for segments within the same city is provided by the city of London, for which the length of segments (expressed in kilometers) spans from 0.00 to 104.97, with a mean of 2.61. By contrast, the range is much smaller (less than 7 km) for the cities of Apeldoorn and Luxembourg in the macro area of Benelux. It is noteworthy that the information on the two cities of Rotterdam and The Hague is here presented separately. However, in the study these urban centers are analysed together to match the definition of the Global Human Settlement - Urban Centers Database \cite{GHSUrbanCentreDatabase}.

\subsection{Remapping of information from Strava segments to streets in the street-network of each city}
\begin{table}[t]
\centering
\caption{OpenStreetMap key-value pairs for the classification of cycleways in protected and unprotected. }
\label{tab:protection_classification}
\begin{tabular}{lll}
\toprule
\textbf{Type} & 
\textbf{OpenStreetMap key} & \textbf{OpenStreetMap value}  \\
Protected cycleway & $highway$ & $['cycleway', 'path', 'footway', 'bridleway', 'track']$ \\
Protected cycleway  & $cycleway$ & $['track', 'opposite_track']$ \\
Protected cycleway  & $cycleway:left$ & $['track', 'opposite_track']$ \\
Protected cycleway  & $cycleway:right$ & $['track', 'opposite_track']$ \\
Protected cycleway  & $cycleway:both$ & $['track', 'opposite_track']$ \\
Protected cycleway  & $bicycle$ & $['designated']$ \\
Unprotected cycleway  & $cycleway$ & $['lane', 'opposite_lane', 'share_busway', 'shared_lane', 'designated', 'yes']$ \\
Unprotected cycleway  & $cycleway:left$ & $['lane', 'opposite_lane', 'share_busway', 'shared_lane', 'designated', 'yes']$ \\
Unprotected cycleway  & $cycleway:right$ & $['lane', 'opposite_lane', 'share_busway', 'shared_lane', 'designated', 'yes']$ \\
Unprotected cycleway  & $cycleway:both$ & $['lane', 'opposite_lane', 'share_busway', 'shared_lane', 'designated', 'yes']$ \\
Unprotected cycleway  & $bicycle$ & $['yes', 'permissive', 'destination', 'private']$ \\
\toprule
\end{tabular}
\end{table}

To identify the gender of cyclists travelling upon the streets network of each city, the collection of Strava segments for each city $c$ was re-projected on the corresponding street-network following the pipeline described below.
\begin{enumerate}
    \item Load the Strava data for city $c$.
    \item Extract the bounding box of city $c$ from the GHS-UCDB R2019A \cite{GHSUrbanCentreDatabase}.
    \item From OpenStreetMap, extract the street-network within the polygon defined in the bounding box using the OSMnx library \cite{boeing2017osmnx}. Set: $network\_type='bike'$, $retain\_all=True$, $simplify=True$.
    \item Classify streets in the street network based on OpenStreetMap attributes in: `street with protected bike lane', `street with unprotected bike lane' and `street with no cycleway'. The $(key,value)$ pairs for the classification are provided in the \tablename~\ref{tab:protection_classification}. All other bikable streets are classified as `no cycleway'.
    \item Proceed with the \textit{preferential assignment} of Strava segments as follows. Buffer with a 10 meters radius the geometries of the street network. Select all streets categorized as `protected cycleways' and intersect each Strava segment with the network. Re-project each segment (or portion(s) of a segment) on all streets with an intersection of at least 30 meters.  Finally compute the geometries of Strava segments left unassigned - that could be either a full segment or portion(s) of a segment- and repeat the procedure selecting `unprotected cycleways' first and subsequently streets with `no cycleways'. 
    \item Compute the gender ratio of each street in the street network using statistics from the re-projected Strava segments. In particular, letting I be the set of segments re-projected to street $s$, $Females_{i}$ ($Males_{i}$) the number of unique female (male) cyclists on segment $i$, the total number of female cyclists on streets $s$ (and correspondingly for male cyclists) is defined as: 
    \begin{equation}
        Females_{s}=\sum_{i\in I} Females_{i}
    \end{equation}
    The gender ratio ($\sigma_{s}$) of street $s$ is then computed as:
    \begin{equation}
        \sigma_{s}=\dfrac{Females_{s}}{Males_{s}+Females_{s}} =\dfrac{\sum_{i\in I} Females_{i}}{\sum_{i\in I} Females_{i}+Males_{i}} 
    \end{equation}
\end{enumerate}
The rationale for the \textit{preferential assignment} is that if a cycleway runs parallel to a street with no cycleway and the linestring geometry for the Strava segment is compatible with both streets (i.e. it falls within the buffered geometry of both streets), we assume that the cyclists rode on the cycleway as opposed to the street with no cycling-dedicated infrastructure.  This approach prevents us from remapping the same portion of a Strava segment to multiple parallel streets with different characteristics.

\subsection{Construction of city-level index of gender gap in recreational cycling}
Strava data remapped on the street-network of each urban area were then used to construct an index of the gender-cycling-gap for all cities included in our sample. The gender-cycling-gap of city $c$ is measured by $\sigma_c$ defined as the ratio between the total kilometers travelled by female cyclists and the overall kilometers travelled by cyclists of both gender within the urban area. The rationale for the use of this metric is its ability to capture two forms of gender gaps described in the literature on cycling and gender, i.e. the propensity of women to make less trips then men and the propensity to cycle shorter distance. This measure is equivalent to the weighted sum of the gender ratio on streets ($\sigma_{s}$) within the urban area, with weights equal to the product of the length and the total popularity (total number of cyclists) of the street. I.e., letting $S$ be the set of streets in the street network of city $C$, $Females_{s}$ ($Males_{s}$) the number of female (male) cyclists on $s$ and $l_{s}$ the length of street $s$ expressed in kilometers:
\begin{equation}
    \sigma_{c}= \dfrac{\sum_{s \in S} Females_{s}*l_{s}}{\sum_{s \in S} (Females_{s}+Males_{s})*l_{s}}=\dfrac{\sum_{s \in S} \sigma_{s}*l_{s}*(Females_{s}+Males_{s})}{\sum_{s \in S}(Females_{s}+Males_{s})*l_{s}}
\end{equation}
A value of $\sigma_{s}$ between 0 and 0.5 indicates the presence of a positive gender gap (with more men cycling then women), while a value above 0.5 indicates a negative gender gap. In our sample, the maximum value of $\sigma_{s}$ stands at 0.2, indicating a positive gender gap for all cities and a monotonic relationship between the gender gap and $\sigma_c$. A full ranking of cities for the four geographical areas is provided in \tablename~\ref{tab:city_ranking}.

\section{Understanding the determinants of gender-cycling-gap - a cross-cities analysis: data, methodology and sensitivity analysis} \label{cross-cities} 
\subsection{Data sources}
A full list of data sources used for this strand of the study is provided below. 
\begin{itemize}
    \item Data on recreational cycling at city-level from Strava. The data were processed following the steps described in the previous section. 
    \item City-indicators from the Global Human Settlement - Urban Centre Database 2015 (GHS-UCDB R2019A) \cite{GHSUrbanCentreDatabase}. The following information was extracted:
    \begin{enumerate}
        \item the share of population living in green areas: data field $SDG\_A2G14$;
        \item the percentage of open space: data field $SDG\_OS15MX$;
        \item the built-up area per capita, data field $BUCAP15$;
        \item the average night-light emission: data field $NTL\_AV$; 
        \item the concentration levels of PM2.5: data field $E\_CPM2\_T14$, 
        \item the city area: data field $AREA$;
        \item the population density: computed as $\dfrac{P15}{AREA}$,
        \item the GDP per person, computed as $\dfrac{GDP15\_SM}{P15}$.
    \end{enumerate} 
    \item Street-network indicators from \cite{boeing2021street}. Out of the list of available indicators, we extract the average absolute street grade, the average degree, orientation order, the proportion of three-way intersections and the average street straightness. 
    \item Urban safety indicators measuring the proportion of the street network with cycleways and the proportion of streets with low speed limit. These data were directly constructed from OpenStreetMap \cite{OpenStreetMap} information following the the pipeline in Section~\ref{sec:road_safety_ind} below.
    \item The Global Gender Gap Index (country-level) from the World Economic Forum \cite{wefgender}. Included in the sensitivity analysis only.
\end{itemize}
It should be noted that the final sample for this component of the analysis consists of 61 cities. The city of New York was excluded from this component of the study due to the large discrepancy between the administrative area of this city and the bounding box of the GHS, which would have made the indicators based on this definition of urban center not representative for the area actually covered by the cycling data. 

\subsection{Construction of urban road safety indicators} \label{sec:road_safety_ind}

OpenStreetMap information accessed via the Python library OSMnx \cite{boeing2017osmnx} was used to construct the two indicators on urban road safety. 
The indicator on the proportion of streets with max-speed limit equal or below 20mi/h or 30km/h (simply referred to as $speed\;limit$ in the Main) was constructed according to the pipeline described below. 
\begin{enumerate}
\item For each city $c$, extract the bounding box of city $c$ from the GHS-UCDB R2019A \cite{GHSUrbanCentreDatabase}.
\item Extract the street-network from the polygon defined in the bounding box using the OSMnx library \cite{boeing2017osmnx}. Set: $network\_type='drive'$, $retain\_all=True$.
\item Compute the proportion of streets satisfying the condition on the speed limit. Weight each street with its length.
\end{enumerate}

The indicator on the proportion of streets with cycling-dedicating infrastructure (simply referred to as $bike\;lanes$) was constructed according to the pipeline below. 
\begin{enumerate}
\item For each city $c$, extract the bounding box of city $c$ from the GHS-UCDB R2019A \cite{GHSUrbanCentreDatabase}.
\item Extract the street-network from the polygon defined in the bounding box using the OSMnx library \cite{boeing2017osmnx}. Set: $network\_type='bike'$, $retain\_all=True$. Call this graph $G0$.
\item From OpenStreetMap \cite{OpenStreetMap}, extract the street-network from the polygon defined in the bounding box using the OSMnx library. Set: $network\_type='drive'$, $retain\_all=True$. Call this graph $G1$.
\item Define as cycleways all streets in $G0$ with the pairs of OSM attribute described in \tablename~\ref{tab:bike_lane_def_index}.
\item Sum over the length of all 'bike-lanes' in $G0$. 
\item Sum over the length of all streets in $G1$. 
\item Define the index as the ratio between the metric computed at point 5 and the metric computed at point 6. 
\end{enumerate}

\begin{table}[t]
\centering
\caption{Definition of bike lanes for construction of city-level indicators of urban road safety. The table provides the OpenStreetMap $(key,value)$ pairs used for the identification of streets with some form of cycling dedicated infrastructure, simply indicated as $bike\;lane$ in the main text. This information was then used to measure the size of the cycling-dedicated-infrastructure in the city and construct the corresponding city-level indicator.}
\label{tab:bike_lane_def_index}
\begin{tabular}{ll}
\toprule
\textbf{OpenStreetMap key} & \textbf{OpenStreetMap value}  \\
 
$highway$ & $['cycleway']$ \\
 $cycleway$ & $['track', 'opposite_track', 'lane', 'opposite_lane', 'opposite', 'share_busway', 'shared_lane', 'designated', 'yes']$ \\
$cycleway:left$ & $['track', 'opposite_track', 'lane', 'opposite_lane', 'opposite', 'share_busway', 'shared_lane', 'designated', 'yes']$ \\
 $cycleway:right$ & $['track', 'opposite_track', 'lane', 'opposite_lane', 'opposite', 'share_busway', 'shared_lane', 'designated', 'yes']$ \\
 $cycleway:both$ & $['track', 'opposite_track', 'lane', 'opposite_lane', 'opposite', 'share_busway', 'shared_lane', 'designated', 'yes']$ \\
\toprule
\end{tabular}
\end{table}

\subsection{Regression analysis} \label{regression}
We estimated a linear regression model via Ordinary Least Squares (OLS) of the form: 
\begin{equation}
    \sigma_c=\sum_{j=1}^{N}\beta_{j}z_{j,c}+\epsilon_{c} \quad  c=1,..,61
\end{equation}
where the list of regressors $z_{j}$ in the preferred model includes: \textit{speed limit}, \textit{orientation}, \textit{GDP},  \textit{3-way crosses}, \textit{night-light emissions}, \textit{grade}, \textit{pm2.5} plus three dummy variables for the macro area to which the city belong (US, UK, Benelux, \textit{baseline}: Italy). All continuous regressors were normalised using a z-score transformation. Out of the initial 15 city-level indicators collected (provided in Table 1) , the final subset of seven indicators (plus the three country-level dummies) included in the regression were selected via exhaustive search to minimize the Akaike Information Criterion (AIC) of the model. The model is estimated using the OLS function of the Python library $statsmodel$ \cite{Seabold2010StatsmodelsEA}.

\subsection{Sensitivity analysis} \label{sensitivity}
\begin{figure}
\centering
\includegraphics[width=0.9\textwidth]{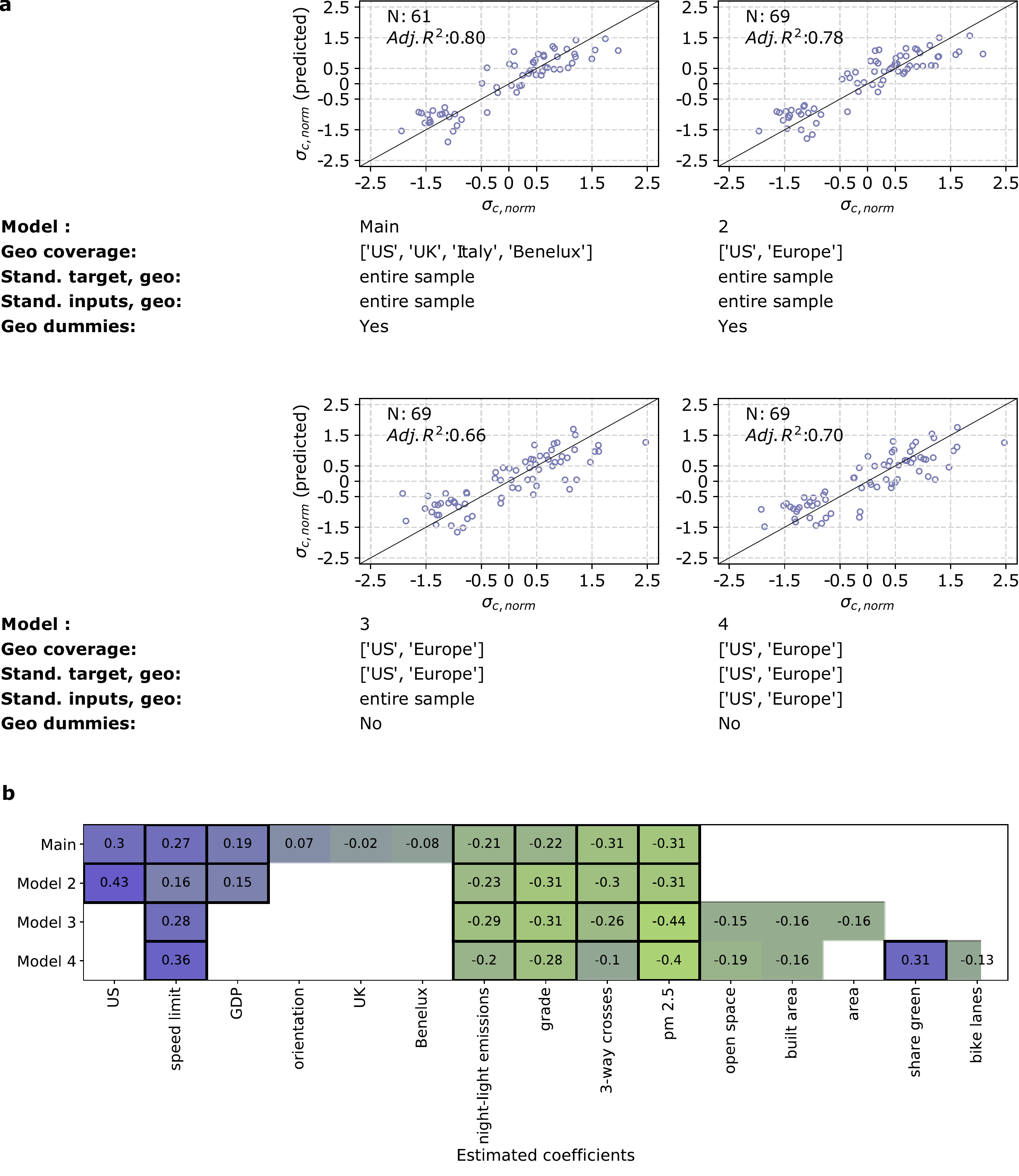}
\caption{\textbf{Understanding the determinants of gender gap in recreational cycling: cross-cities comparison: sensitivity analysis.} The panel provides a comparison between the preferred model (Main model) discussed in the main text and three additional models, included as sensitivity analysis. a) A summary of the characteristics of each model and scatter plots of (normalized) actual vs fitted values, for each model separately. All models share the linear formulation and the estimation technique (OLS) with the main one, but differ in terms of geographical coverage, standardization sample for target and input variables and the presence/absence of geographical dummies (Section \ref{cross-cities}\ref{sensitivity}). b) The heatmap displaying the estimated coefficient for each model. Cells outlined in black correspond to coefficients statistically significant at 0.05 significance level. }
\label{fig:sensitivity}
\end{figure}

As sensitivity analysis, we estimated three additional regression models, in which we adopted different strategies to account for the heterogeneous levels of penetration of Strava across countries covered in the study. The three additional models share the linear formulation and the estimation technique (OLS) with the preferred one, but differ in (at least one) of the following characteristics:
\begin{enumerate}
    \item Standardization of target and input variables on either the full sample or by geographical area. For model 2, the standardization was still performed on the full sample (as for the preferred model) for both target and input variables. For model 3, the target variable was standardized by continent ('US' vs 'Europe') while the input variables were standardized at the level of the entire sample. For model 4, both target and input variables were standardized by continent ('US' vs 'Europe').  
    \item Inclusion and exclusion of geographical dummies. Model 2 includes a geographical dummy for the US (with baseline: Europe). Model 3 and 4 do not include geographical dummies, but the standardization of the target and/or inputs is performed by geographical area ('US' vs 'Europe').
    \item Enlargement of the sample to include 8 additional cities in Europe, located outside of the four main geographical areas covered by the study.
\end{enumerate}
A characterization of each model is provided in \figurename~\ref{fig:sensitivity}-a. Model 1 corresponds to the preferred model described in the main analysis.
Variables selection for each additional model was performed via extensive search to minimize the Akaike Information Criterion (AIC) of the model.
Scatter plots of the actual values vs the fitted values of $\sigma_c$ for each model are presented in \figurename~\ref{fig:sensitivity}-a, suggesting that overall all four models explain well the variability in the observed $\sigma_c$. Figure~\ref{fig:sensitivity}-b shows the estimated coefficients of the preferred model and the three additional models. Matrix cells outlined in black indicate that the corresponding coefficient is statistically significant at 0.05 level. The sensitivity analysis highlights that the results discussed in the main text concerning the role of the speed limit, average hilliness and complexity of the street network are robust across the different specifications, both in terms of the sign of the effect of each regressor and its significance.

\section{Case study on the City of New York: data and methodology}

\subsection{Data sources}
A full list of data sources used for this component of the study is provided below. 
\begin{itemize}
    \item Data on recreational cycling at street-level for the city of New York from Strava. The raw Strava data were processed and remapped to the street-network of each city extracted from OpenStreetMap following the steps described in Section \ref{sec:StravaData}. A network definition of streets was used, which does not reflect a the toponymy of streets.
    \item OpenStreetMap data on street-level characteristics extracted during the process of remapping of Strava data via the python library OSMnx \cite{boeing2017osmnx}. In particular, for each street, we retained information on: the presence of public lighting, the presence of protected or unprotected cycleways, proximity with a park or with the coastline and whether the surface is paved. A list of OSM key-value pairs is provided in \tablename~\ref{tab:Street-level features}.
    In addition, for streets in the largest component of the street network, we computed the edge-betweenness \cite{latora2017complex} via the Python library \textit{graph-tool} \cite{peixoto_graph-tool_2014}. Streets outside the largest component of the network (i.e. excluding streets in the borough of Staten Island) were excluded from the sample.
    \item Administrative data from the OpenData Portal of the city of New York on location of all (any-vehicle) accidents and bike accidents only. The data of car and bike accidents were processed to compute the number of car and bike accidents per 10 meters, for each street. The raw data are available at \cite{NYCadmin}.
    \item Shapefiles of the administrative boundaries of boroughs in the city of New York. Available at \cite{NYCadmin}.
\end{itemize}

\clearpage

\begin{turnpage}

\begin{table}[t!]
\centering
\caption{List of street-level urban features included in the New York City case study.\footnote{*Indicates that the data from the original data sources required specific processing (e.g. for normalisation).}}
\label{tab:Street-level features}
\begin{tabular}{p{0.12\linewidth} | p{0.25\linewidth} |p{0.05\linewidth} |p{0.58\linewidth}}
\toprule
\textbf{Variable name} & \textbf{Description} &
  \textbf{Data source} &
  \textbf{OSM key-values pairs} \\ 
Unprotected cycleway & Dummy for presence of a shared or unprotected bike-lane &
  OSM &
  \begin{tabular}[c]{@{}l@{}}\{'cycleway:right': {[}'lane', 'opposite\_lane', 'share\_busway', 'shared\_lane', 'designated', 'yes'{]}, \\ 'cycleway:left': {[}'lane', 'opposite\_lane', 'share\_busway', 'shared\_lane', 'designated', 'yes'{]}, \\ 'cycleway:both': {[}'lane', 'opposite\_lane', 'share\_busway', 'shared\_lane', 'designated', 'yes'{]}, '\\ cycleway': {[}'lane', 'opposite\_lane', 'opposite', 'share\_busway', 'shared\_lane', 'designated', 'yes'{]}, \\ 'bicycle': {[}'yes', 'permissive', 'destination', 'private'{]}\}\end{tabular} \\
Protected cycleway & Dummy for either the presence of a protected bike-lane or streets with no vehicles &
  OSM &
  \begin{tabular}[c]{@{}l@{}}\{'highway': {[}'cycleway', 'path', 'footway', 'bridleway', 'track'{]}, \\ 'cycleway': {[}'track', 'opposite\_track'{]}, \\ 'cycleway:left': {[}'track', 'opposite\_track'{]}, \\ 'cycleway:right': {[}'track', 'opposite\_track'{]}, \\ 'cycleway:both': {[}'track', 'opposite\_track'{]}, \\ 'bicycle': {[}'designated'{]}\} \end{tabular} \\
Public lighting &Dummy for the presence of public lighting &
  OSM &
  \{'lit':{[}'yes', '24/7'{]}\} or missing tag \\
Unpaved surface &Dummy for unpaved surface &
  OSM &
  \{'surface':{[}'unpaved', 'compacted', 'fine\_gravel', ''gravel', 'pebblestone', 'ground', 'earth', 'dirt','grass', 'grass\_paver', 'sand', 'mud'{]}\} \\
Park proximity & Dummy for streets next to a park (within 15 meters) & 
  OSM* &
  \{'leisure':'park':\} + 15m proximity from the geometry \\
Any-vehicle crashes & Number of crashes involving any type of vehicles per 10m of street length &
  NYC* &
  Not applicable \\
Bike crashes & Number of bike crashes per 10m of street length &
  NYC* &
  Not applicable \\
\{borough\_name\} & Dummy for boroughs (baseline: Manhattan)&
  NYC &
  Not applicable \\
Coast proximity & Dummy for street next to the river coast&
  OSM* &
  {'natural':'coastline'} + 150m proximity from the geometry  \\
Edge betweenness & Edge betweenness of the streets computed for streets within the largest component of the street network &
  OSM* &
  OSM street-network used to compute edge-betweenness on the largest component of the network. Network extracted using OSMNX, network\_type='bike'. \\ \toprule
\end{tabular}
\end{table}

\end{turnpage}

\subsection{Methodology}

\begin{figure}
\includegraphics[width=0.9\textwidth]{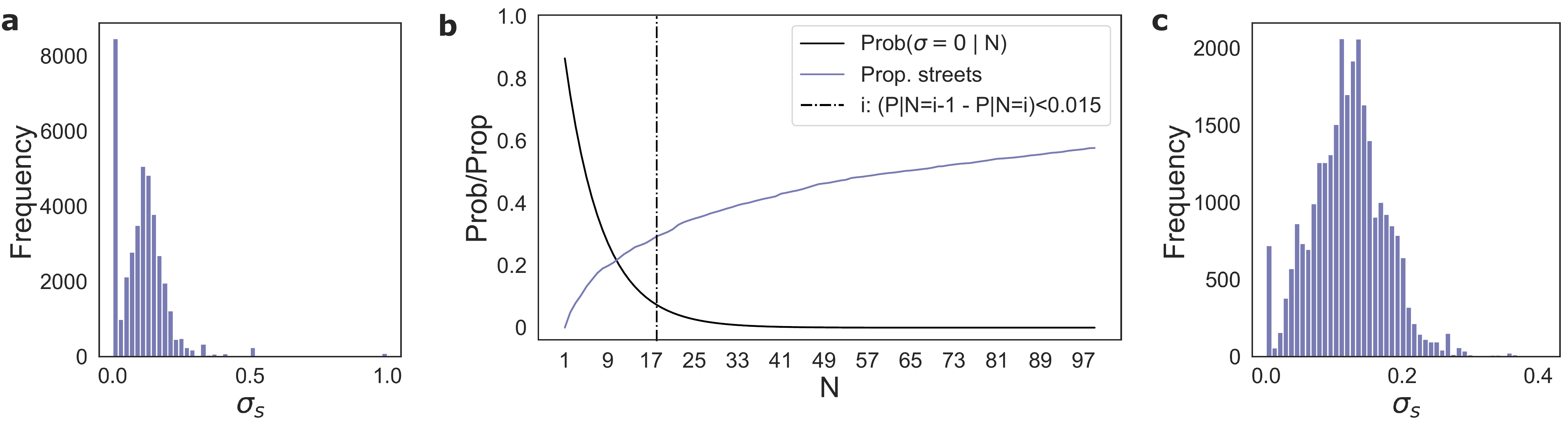}
\caption{\textbf{Impact of streets filtering}. a) The distribution of $\sigma_s$ for streets in the City of New York, before filtering. b) The black line depicts the probability of observing $\sigma_s=0$ conditional on the number of cyclists. The light blue line depicts the proportion of streets in the street network as a function of the number of cyclists. The vertical dashed line depicts the selected threshold for the data filtering. c) The distribution of $\sigma_s$ for streets in the City of New York, after filtering of segments with number of cyclists below the selected threshold.}
\label{fig:datafiltering}
\end{figure}

\paragraph*{Data filtering}
The vast majority of Strava users are men. Because of this, the probability of observing $\sigma_s=0$ is a decreasing function of the number of cyclists on the street. As such, observing no women on a road may be due to two factors. First, it may be the case that the street is not (overall) sufficiently popular: being women underrepresented among Strava users, there will not be female cyclists on it. On the other hand, the segment might be overall sufficiently popular, but it's attractiveness being low among women compared to men. These effects are hard to disentangle, limiting our ability to interpret an extreme values of $\sigma_s$, on segments with low overall popularity.
To mitigate this issue, we filter the data to exclude those segments with a small number of cyclists.
In particular, for a city $c$, the probability of observing $\sigma_{s}=0$ on  a segment $s$ conditional to observing $N_s$ cyclists on $s$ is given by (assuming replacement): 
\begin{equation}
    P_{N}=P(\sigma_{s}=0 | N_{s}) = \bigg(\dfrac{Males_{c}}{Males_{c}+Females_{c}}\bigg)^{N_{s}}
\end{equation}
which is decreasing in $N_{s}$.
To ensure that there is a weak dependence between the popularity of the segment and the observed gender ratio, we choose a filtering threshold on $N_s$ such that we retain only those segments $s$ for which  $P_{N_{s}}-P_{N_{s}-1}$ is low( we fixed $P_{N_{s}}-P_{N_{s}-1}<0.015$). The impact of this filtering-rule on the distribution of gender ratio for the city of New York is depicted in \figurename~\ref{fig:datafiltering}. The top panel shows $P_{N_{s}}$ as a function of $N_s$ for the City of New York, together with the vertical line which denotes the selected threshold according to the aforementioned criterion. The bottom panels show the distribution of gender ratio before (left) and after (right) the filtering. As expected, the observed zero inflation of the distribution is much less marked after the filtering, confirming the effectiveness of the adopted mitigation strategy. In addition, the new distribution appears much less right-skewed.

\paragraph*{Multivariate logistic regression}
To assess the degree of association between $\sigma_s$ and the presence of cycling-dedicated infrastructure, we use a multivariate logistic regression. We restrict the sample to streets belonging to the bottom and top 33\% of the distribution fo $\sigma_s$ and classify streets in $Low$ and $High$ $\sigma_s$ respectively. As a robustness check, the analysis is repeated for alternative values of this threshold (0.25 and 0.40, instead of 0.33). We use features described in Table \ref{tab:Street-level features} as predictors and the binarized $\sigma_s$ as the target variable. Moreover, we
scale continuous predictors (\textit{Any-vehicle crashes}, \textit{Bike crashes} and \textit{Edge-betweenness}) using a z-score-transformation to normalize the magnitude of the estimated coefficients.
For estimation of the Multivariate Logistic regression, we use the Logit function of the Python library $statsmodel$ \cite{Seabold2010StatsmodelsEA}.

\newpage

\begin{table*}
\caption{Geographic characteristics of raw Strava segments by city. For each city, the table reports the total number of Strava segments, the length of the shortest segment in the data collection in kilometers (Min (km)), the length of the longest segment in the data collection in kilometers (Max (km)), the mean value of the distribution of the lengths of segments (Mean (km)) and its standard deviation (Std (km)), expressed in kilometers. All figures are rounded to the nearest 0.01.}
\centering
\footnotesize
\begin{tabular}{llllllll}
\toprule
& \textbf{City name}               & \textbf{Country} & \textbf{N seg.} & \textbf{Min (km)}  & \textbf{Max (km)}    & \textbf{Mean (km)} & \textbf{Std (km) }  \\
 
0  & Albuquerque         & USA     & 1997        & 0.25 & 126.51 & 6.88    & 8.97  \\
1  & Almere              & Benelux & 818         & 0.08 & 24.03  & 2.31    & 3.13  \\
2  & Amsterdam           & Benelux & 2091        & 0.00  & 23.39  & 1.64    & 2.13  \\
3  & Antwerp             & Benelux & 1014        & 0.04 & 23.96  & 1.67    & 2.68  \\
4  & Apeldoorn           & Benelux & 357         & 0.03 & 7.12   & 1.14    & 1.04  \\
5  & Arnhem              & Benelux & 1608        & 0.00  & 19.75  & 1.24    & 1.79  \\
6  & Austin              & USA     & 5310        & 0.07 & 143.23 & 7.16    & 12.00  \\
7  & Bari                & Italy   & 129         & 0.03 & 14.50   & 2.25    & 2.63  \\
8  & Bologna             & Italy   & 653         & 0.10  & 44.52  & 2.85    & 4.68  \\
9 & Boston              & USA     & 1400        & 0.19 & 53.87  & 4.45    & 5.96  \\
10 & Breda               & Benelux & 457         & 0.07 & 18.42  & 1.64    & 2.24  \\
11 & Brescia             & Italy   & 665         & 0.03 & 21.19  & 2.20     & 2.76  \\
12 & Bristol             & Uk      & 3221        & 0.00  & 40.34  & 1.61    & 2.87  \\
13 & Catania             & Italy   & 271         & 0.14 & 49.78  & 3.95    & 6.40   \\
14 & Charleroi           & Benelux & 355         & 0.07 & 13.51  & 1.12    & 1.53  \\
15 & Charlotte           & USA     & 1664        & 0.05 & 136.31 & 7.50     & 13.18 \\
16 & Derby               & UK      & 1381        & 0.04 & 18.27  & 1.34    & 2.04  \\
17 & Eindhoven           & Benelux & 779         & 0.03 & 18.11  & 1.27    & 1.57  \\
18 & Exeter              & UK      & 1674        & 0.00  & 24.83  & 1.70     & 2.94  \\
19 & Florence            & Italy   & 689         & 0.00  & 24.50   & 1.53    & 1.82  \\
20 & Genoa               & Italy   & 2107        & 0.00  & 35.92  & 2.35    & 2.88  \\
21 & Ghent               & Benelux & 1170        & 0.04 & 20.71  & 1.26    & 1.97  \\
22 & Groningen           & Benelux & 1279        & 0.00  & 11.34  & 1.45    & 1.51  \\
23 & Haarlem             & Benelux & 220         & 0.01 & 10.78  & 1.11    & 1.29  \\
24 & Jacksonville        & USA     & 1032        & 0.36 & 160.42 & 10.54   & 16.88 \\
25 & Las Vegas            & USA     & 1948        & 0.01 & 102.45 & 7.63    & 9.40   \\
26 & Leeds               & UK      & 9991        & 0.00  & 82.89  & 2.90     & 5.87  \\
27 & Liege               & Benelux & 770         & 0.08 & 14.62  & 1.19    & 1.20   \\
28 & London              & UK      & 18232       & 0.00  & 104.97 & 2.61    & 4.69  \\
29 & Louisville          & USA     & 1833        & 0.19 & 208.30  & 9.15    & 14.50  \\
30 & Luxembourg          & Benelux & 606         & 0.06 & 7.36   & 1.14    & 0.93  \\
31 & Manchester          & UK      & 2812        & 0.02 & 23.17  & 1.65    & 2.47  \\
32 & Memphis             & USA     & 709         & 0.06 & 102.86 & 9.68    & 14.41 \\
33 & Modena              & Italy   & 191         & 0.13 & 17.24  & 3.27    & 3.52  \\
34 & Nashville           & USA     & 2186        & 0.13 & 168.36 & 7.78    & 12.32 \\
35 & Newcastle upon Tyne & UK      & 2085        & 0.03 & 17.23  & 1.51    & 2.21  \\
36 & New York             & USA     & 7122        & 0.00  & 256.20  & 11.13   & 19.44 \\
37 & Nijmegen            & Benelux & 551         & 0.10  & 5.75   & 0.93    & 0.86  \\
38 & Norwich             & UK      & 1065        & 0.00  & 20.73  & 1.14    & 1.82  \\
39 & Nottingham          & UK      & 2168        & 0.02 & 22.73  & 1.43    & 2.30   \\
40 & Oklahomacity        & USA     & 1502        & 0.23 & 277.57 & 16.75   & 28.65 \\
41 & Padua               & Italy   & 130         & 0.20  & 10.47  & 1.58    & 1.40   \\
42 & Palermo             & Italy   & 929         & 0.10  & 57.93  & 3.15    & 4.23  \\
43 & Parma               & Italy   & 155         & 0.18 & 17.93  & 3.77    & 3.23  \\
44 & Phoenix             & USA     & 6564        & 0.05 & 219.64 & 11.22   & 17.39 \\
45 & Plymouth            & UK      & 3503        & 0.03 & 47.29  & 1.63    & 2.70   \\
46 & Prato               & Italy   & 288         & 0.17 & 15.19  & 1.94    & 2.02  \\
47 & Reading             & UK      & 830         & 0.08 & 16.75  & 1.09    & 1.51  \\
48 & Reggio Emilia        & Italy   & 178         & 0.06 & 28.69  & 3.55    & 3.91  \\
49 & Rome                & Italy   & 3257        & 0.00  & 52.60   & 3.00     & 5.11  \\
50 & Rotterdam           & Benelux & 1871        & 0.00  & 20.35  & 1.76    & 2.42  \\
51 & San Antonio          & USA     & 3277        & 0.11 & 167.88 & 8.68    & 14.84 \\
52 & Sheffield           & UK      & 7733        & 0.00  & 88.06  & 2.50     & 5.11  \\
53 & Southampton         & UK      & 1495        & 0.07 & 17.03  & 1.34    & 1.68  \\
54 & Taranto             & Italy   & 152         & 0.07 & 11.87  & 1.95    & 2.25  \\
55 & The Hague            & Benelux & 1018        & 0.02 & 23.95  & 1.27    & 1.90   \\
56 & Tilburg             & Benelux & 525         & 0.05 & 24.37  & 1.62    & 2.41  \\
57 & Trieste             & Italy   & 1092        & 0.00  & 19.19  & 1.92    & 2.22  \\
58 & Turin               & Italy   & 1418        & 0.00  & 19.27  & 2.06    & 2.10   \\
59 & Utrecht             & Benelux & 1221        & 0.06 & 50.08  & 2.05    & 4.15  \\
60 & Venice              & Italy   & 341         & 0.15 & 75.69  & 3.31    & 7.39  \\
61 & Verona              & Italy   & 1154        & 0.07 & 29.3   & 2.19    & 2.43  \\
62 & York                & UK      & 1711        & 0.03 & 52.88  & 2.12    & 3.45 \\
\toprule
\end{tabular}
\label{tab:Strava_raw_descr_a}
\end{table*}

\newpage

\begin{table}
\centering
\caption{Ranking of cities by $\sigma_c$, by geographical area. The table reports the ranking of cities for the four geographical areas. All figures are rounded to the nearest 0.01.}
\footnotesize
\begin{tabular}{lllll}
\toprule
\textbf{Ranking}& \textbf{City name}               & \textbf{Geographical area} & \textbf{Country} & \textbf{$\sigma_{c}$}  \\
 
1 &	Groningen	& Benelux &	Netherlands	& 0.21 \\
2 &	Utrecht	& Benelux & 	Netherlands	 & 0.17 \\
3  &	Amsterdam &	Benelux &	Netherlands &	0.16\\
4  &	Apeldoorn &	Benelux &	Netherlands &	0.15 \\
5 &	Nijmegen &	Benelux	 &Netherlands &	0.15 \\
6 &	Haarlem &	Benelux &	Netherlands &	0.14 \\
7 &	Almere &	Benelux &	Netherlands &	0.13\\
8 &	Eindhoven &	Benelux &	Netherlands &	0.13 \\
9 &	Arnhem &	Benelux &	Netherlands &	0.13 \\
10 &	Breda &	Benelux &	Netherlands &	0.13 \\
11 &	Tilburg &	Benelux &	Netherlands &	0.12 \\
12 &	Rotterdam/The Hague &	Benelux &	Netherlands &	0.12 \\
13 &	Ghent &	Benelux &	Belgium &	0.10\\
14 &	Luxembourg city &	Benelux &	Luxembourg &	0.09\\
15 &	Antwerp &	Benelux &	Belgium &	0.09\\
16 &	Liege &	Benelux &	Belgium &	0.06 \\
17 &	Charleroi &	Benelux &	Belgium &	0.06 \\
 
1 &	Venice &	Italy &	Italy &	0.09 \\
2 &	Padua &	Italy &	Italy &	0.07\\
3 &	Verona &	Italy &	Italy &	0.07 \\
4 &	Trieste &	Italy &	Italy &	0.06 \\
5 &	Florence &	Italy &	Italy &	0.06 \\
6 &	Parma &	Italy &	Italy &	0.06 \\
7 &	Turin &	Italy &	Italy &	0.05 \\
8 &	Palermo &	Italy &	Italy &	0.05 \\
9 &	Modena &	Italy &	Italy &	0.05 \\
10 &	Bologna	 &Italy &	Italy &	0.04 \\
11 &	Bari &	Italy &	Italy &	0.04 \\
12 &	Brescia &	Italy &	Italy &	0.04 \\
13 &	Genoa &	Italy &	Italy &	0.04 \\
14 &	Reggio Emilia &	Italy &	Italy &	0.04 \\
15 &	Prato &	Italy &	Italy &	0.04 \\
16 &	Catania &	Italy &	Italy &	0.03 \\
17 &	Rome &	Italy &	Italy &	0.03 \\
18 &	Taranto &	Italy &	Italy &	0.02\\
 
1 &	Exeter &	UK &	UK &	0.17 \\
2 &	York &	UK &	UK &	0.15 \\
3 &	London &	UK &	UK &	0.14 \\
4 &	Bristol &	UK &	UK &	0.14 \\
5 &	Nottingham &	UK &	UK &	0.14 \\
6 &	Norwich &	UK &	UK &	0.14\\
7 &	Newcastle upon Tyne &	UK &	UK &	0.13 \\
8 &	Manchester &	UK &	UK &	0.13 \\
9 &	Derby &	UK &	UK &	0.12 \\
10 &	Leeds &	UK &	UK &	0.12 \\
11 &	Southampton &	UK &	UK &	0.12 \\
12 &	Plymouth &	UK &	UK &	0.11 \\
13 &	Reading &	UK &	UK &	0.11 \\
14 &	Sheffield &	UK &	UK &	0.10 \\
 
1 &	Albuquerque &	US &	US &	0.19\\
2 &	Oklahomacity &	US &	US &	0.19 \\
3 &	Boston &	US &	US &	0.18 \\
4 &	Jacksonville &	US &	US &	0.17 \\
5 &	Las Vegas &	US &	US &	0.17 \\
6 &	San Antonio &	US &	US &	0.16 \\
7 &	Nashville &	US &	US &	0.15 \\
8 &	Austin &	US &	US &	0.15 \\
9 &	Memphis	 &US &	US &	0.14 \\
10 &	Louisville &	US &	US &	0.14\\
11 &	Charlotte &	US &	US &	0.13\\
12 &	Phoenix &	US &	US &	0.11\\
\toprule
\end{tabular}
\label{tab:city_ranking}
\end{table}

\newpage

\newpage
\begin{figure}
\centering
\includegraphics[width=0.5\textwidth]{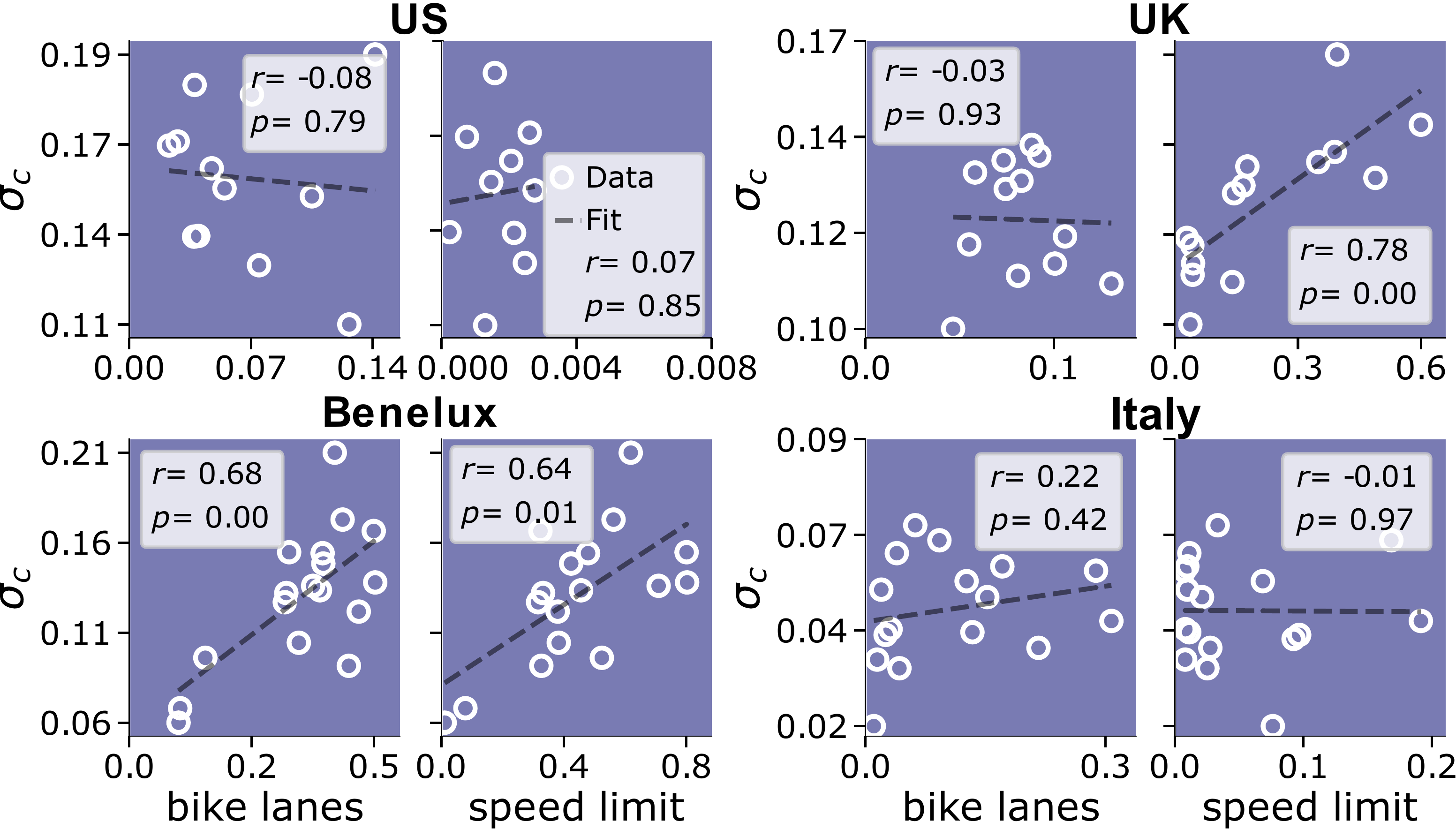}
\caption{\textbf{Correlations between gender ratio and urban road safety indicators, excluding outliers.} The scatter plots show the correlations between two urban road safety indicators and the gender ratio $\sigma_c$, for cities in the four geographical areas separately. For each area, outliers to the three distributions of $\sigma_c$, $bike\;lanes$ and $speed\;limit$ were identified using the IQR Score method and excluded. Each data point represents a city. The black line is the linear fit. The two urban road safety indicators capture the density of streets with a bike-lane in the street-network (indicator: $bike\;lanes$) and the density of streets with a speed limit up to 20 mi/h or 30 km/h in the street network (indicator: $speed\;limit$).}
\label{fig:correlations_noout}
\end{figure}

\begin{figure}
\centering
\includegraphics[width=0.6\textwidth]{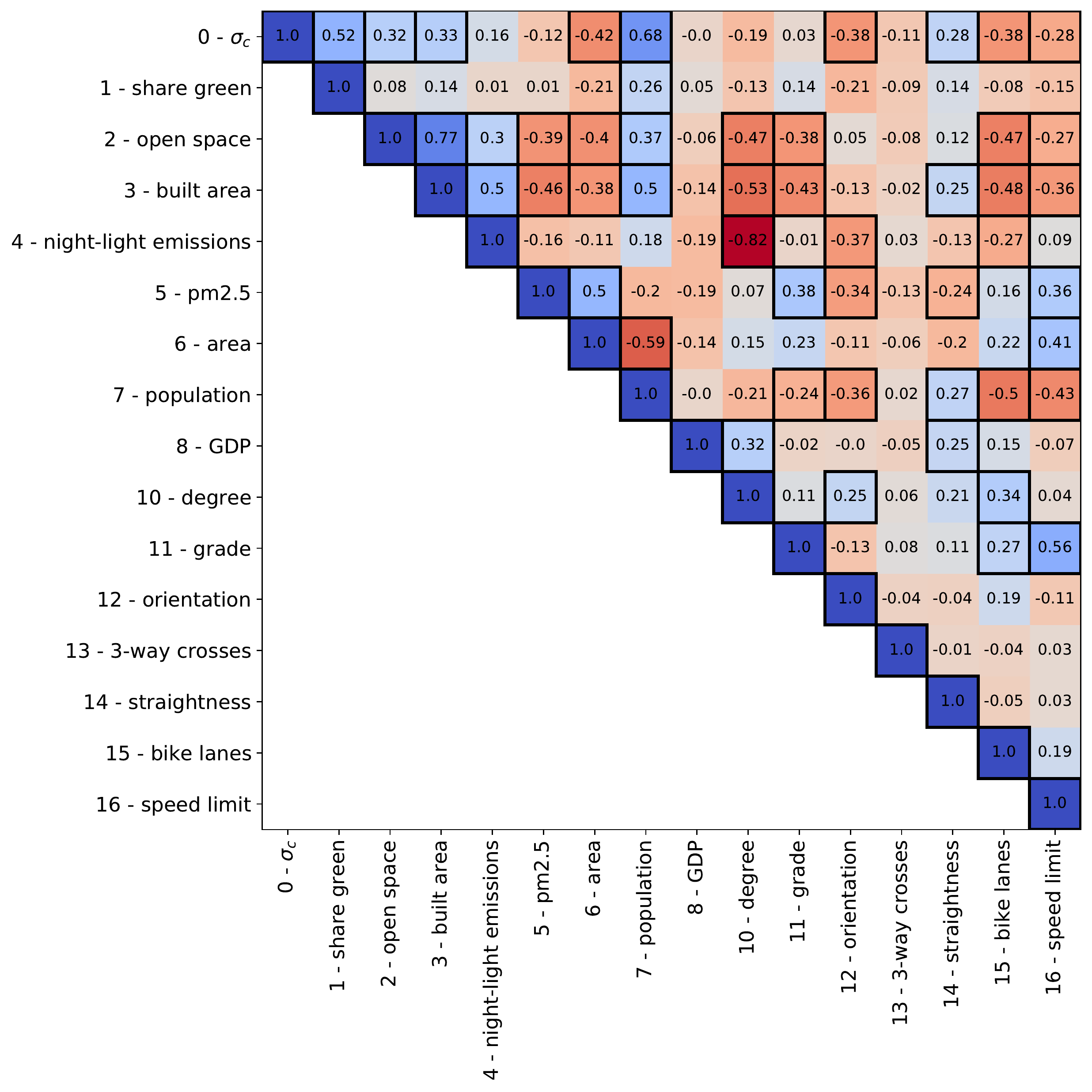}
\caption{\textbf{Correlations between gender ratio and city-level features.} Correlation matrix among city-level characteristics. The correlations are computed for the entire sample of 61 cities on z-scored transformed variables. Cells outlined in black indicate that the correlation is statistically different from 0 at a significance level of $\alpha=0.05$.}
\label{fig:correlations_all}
\end{figure}

\newpage
\begin{figure}
\includegraphics[width=0.9\textwidth]{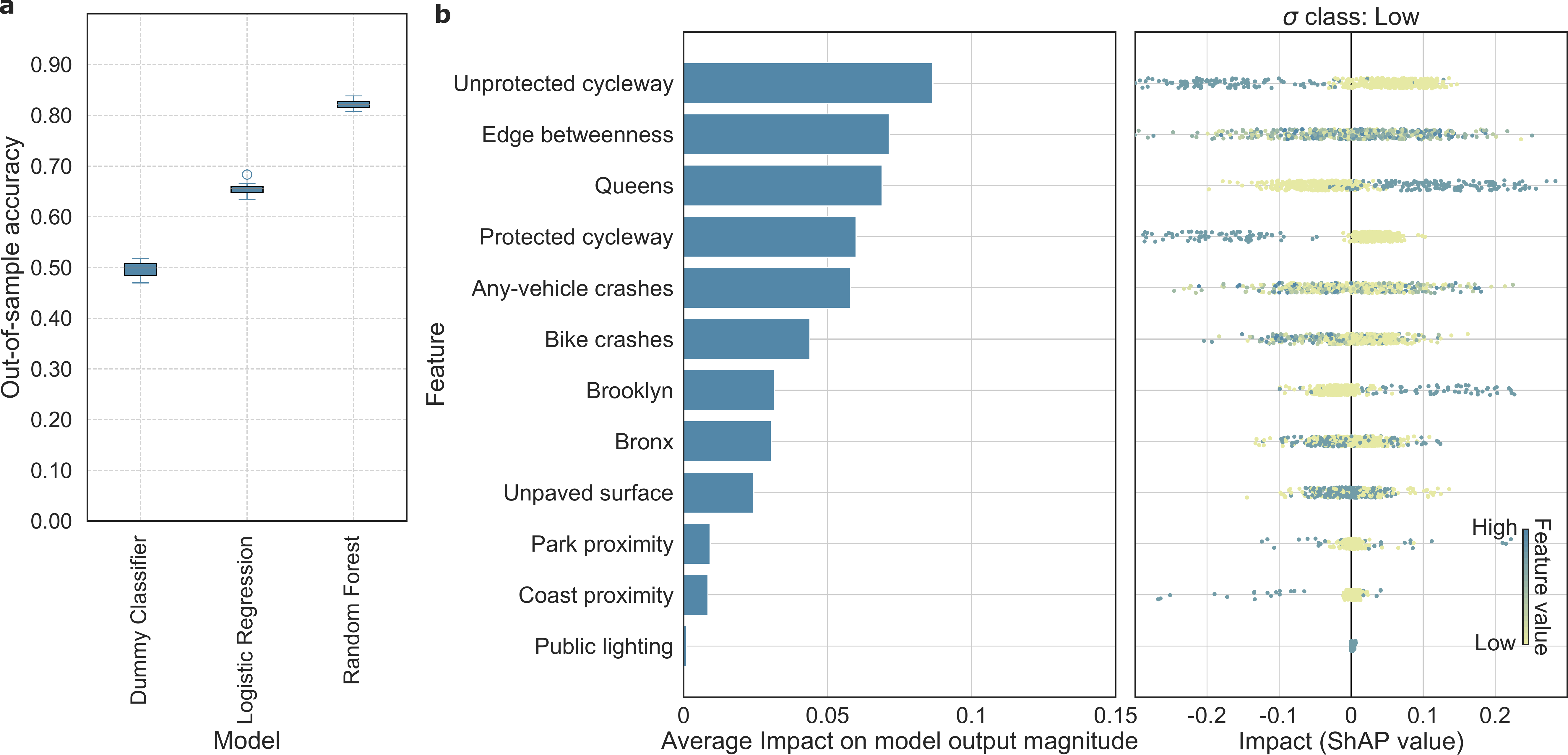}
\caption{\textbf{Classification task, city of New York - models comparison.}  a) The boxplot present the out-of-sample accuracy of the logistic regression vs random-forest, for a value of the threshold $\alpha=0.33$, computing using a stratified 10-fold approach.  b)  The bar plot presents average shapely values computed on a random selection of 500 data point. The adjacent strip plot presents the impact (in terms of shapley values) of a given feature on the probability that the street belong to the $Low$ class, for 500 randomly selected data points \cite{lundberg2017unified}. Each point on the summary plot is a shapley value for a feature and an observation. The color represents the value of the feature from low to high. The negative values associated with the mass of blue dots for unprotected and protected cycleways indicate that, in the random selection of points here presented, the presence of a cycleway decrease the probability that the streets belong to the $Low$ $\sigma_s$ class. Given the symmetry of shape values in a binary classification task, this additionally indicates that for the data points in  random selection, the presence of a cycleway increases the probability that the streets belongs to the $high$ $\sigma_s$ class. }
\label{fig:MLplusShap}
\end{figure}

\end{document}